\documentclass[twocolumn,notitlepage,prl,superscriptaddress,longbibliography]{revtex4-2}

\usepackage{amsfonts}
\usepackage{textcomp}
\usepackage{times}
\usepackage{graphicx}
\usepackage{float}
\usepackage{latexsym,amsmath,amssymb,bm,euscript}
\usepackage{color}
\usepackage{subfigure}
\usepackage{epstopdf}
\usepackage[colorlinks=true,linkcolor=blue,citecolor=blue]{hyperref}
\usepackage{soul}
\usepackage[normalem]{ulem}
\usepackage{mathrsfs}
\usepackage{amsmath}
\usepackage{xspace}
\usepackage{natbib}
\usepackage{ulem}
\usepackage{etoolbox}
\usepackage{tikz}
\usepackage{physics}
\usepackage{chemformula}
\usepackage{natbib}
\usepackage{tikz}

\newcommand{\nbcp}{\ch{Na$_2$BaCo(PO$_4$)$_2$}}

\graphicspath{{./Fig/}}

\begin{document}
\title{{Double Magnon-Roton Excitations in the Triangular-Lattice Spin Supersolid}}

\author{Yuan Gao}
\thanks{These authors contributed equally to this work.}
\affiliation{School of Physics, Beihang University, Beijing 100191, China}
\affiliation{Institute of Theoretical Physics, Chinese Academy of Sciences, 
Beijing 100190, China}
		
\author{Chuandi Zhang}
\thanks{These authors contributed equally to this work.}
\affiliation{School of Physics, Beihang University, Beijing 100191, China}

\author{Junsen Xiang}
\affiliation{Beijing National Laboratory for Condensed Matter Physics, 
Institute of Physics, Chinese Academy of Sciences, Beijing 100190, China}

\author{Dehong Yu}
\affiliation{Australian Nuclear Science and Technology Organisation, 
Lucas Heights NSW 2234, Australia}

\author{Xingye Lu}
\affiliation{Center for Advanced Quantum Studies and Department of Physics,
Beijing Normal University, Beijing 100875, China}

\author{Peijie Sun}
\affiliation{Beijing National Laboratory for Condensed Matter Physics, 
Institute of Physics, Chinese Academy of Sciences, Beijing 100190, China}

\author{Wentao Jin}
\email{wtjin@buaa.edu.cn}
\affiliation{School of Physics, Beihang University, Beijing 100191, China}

\author{Gang Su}
\email{gsu@ucas.ac.cn}
\affiliation{Institute of Theoretical Physics, Chinese Academy of Sciences, 
Beijing 100190, China}
\affiliation{Kavli Institute for Theoretical Sciences, University of Chinese 
Academy of Sciences, Beijing 100190, China}

\author{Wei Li}
\email{w.li@itp.ac.cn}
\affiliation{Institute of Theoretical Physics, Chinese Academy of Sciences, 
Beijing 100190, China}
\affiliation{Peng Huanwu Collaborative Center for Research and Education,
Beihang University, Beijing 100191, China}

\date{\today}

\begin{abstract}
Supersolid is an exotic quantum state of matter that spontaneously hosts the features 
of both solid and superfluid, which breaks the translation and U(1) gauge symmetries. 
Here we {study the spin dynamics in the} triangular-lattice compound \nbcp, which 
is revealed in [Xiang \textit{et al.}, Nature 625, 270-275 (2024)] as a quantum magnetic 
analog of supersolid. {We simulate the easy-axis Heisenberg model with tensor 
network approach and uncover unique dynamic traits. These features are manifested in 
two branches of excitations that can be associated with the spin solidity and superfluidity, 
respectively. One branch contains the U(1) Goldstone and roton modes, while the other 
comprises pseudo-Goldstone and roton modes. The gapless Goldstone modes of the 
in-plane superfluid order are confirmed by our inelastic neutron scattering measurements. 
Together with the evident out-of-plane solid order {indicated} by the magnetic Bragg 
peaks, our findings provide spectroscopic evidence for spin supersolidity in this easy-axis 
antiferromagnet. Akin to the role of phonon-roton modes --- Landau elementary 
excitations ---} in shaping the helium superfluid thermodynamics, the intriguing {double 
magnon-roton dispersion here determines} the low-temperature thermodynamics of 
spin {supersolids} down to sub-Kelvin regime, explaining the recently observed
giant magnetocaloric effect in \nbcp.
\end{abstract}

\maketitle

%%%%%%%%%%%%%%
{\textit{Introduction.---}} 
As a paradigmatic frustrated quantum spin system, the {spin-1/2} triangular-lattice 
antiferromagnets (TLAFs) have garnered significant research interest in the past
\cite{Chubukov1991,Collins1997,Starykh2015}. Compounds with perfect 
{equilateral} triangles, including the cobaltate Ba$_3$CoSb$_2$O$_9$
\cite{Doi2004,Shirata2012,Susuki2013,Zhou2012,Ma2016Dynam,Ito2017Dynam,
Macdougal2020}, rare-earth compounds such as REMgGaO$_4$ (with RE = rare 
earth)~\cite{Li2015YMGO1,Li2015YMGO2,Shen2016,Paddison2017,Shen2019,
Li2020TMGO,Li2020,Hu2020TMGO} and structurally similar compounds ARECh$_2$ 
with A = Na, K, Cs, and Ch = O, S, Se~\cite{Liu2018,Bordelon2019,Bordelon2020,
Ranjith2019,Guo2020,Zhang2021NaYbSe,Dai2021,Scheie2024NP}, have been 
synthesized and investigated recently. {There are studies on the exotic spin states, 
including the unconventional magnetic orders~\cite{Yamamoto2014,Yamamoto2015,
Sellmann2015,Gao2022Super} and possible quantum spin liquids (QSLs)
\cite{Anderson1973,Balents2010,Zhou2017,Broholm2020}, etc., as well as their 
spin excitations~\cite{Zheng2006PRL,Zheng2006,Verresen2019Avoided,Shen2016,
Paddison2017,Kamiya2018,Dai2021,Chi2022,Scheie2024NP} and anomalous 
thermodynamics~\cite{Elstner1993,Chen2019,Li2020TMGO,Hu2020TMGO}. }
The TLAFs provide a very intriguing and fertile ground for emergent quantum spin 
states and phenomena.

% ========== Fig1 ========= %
\begin{figure}[!htbp]
\includegraphics[angle=0,width=1\linewidth]{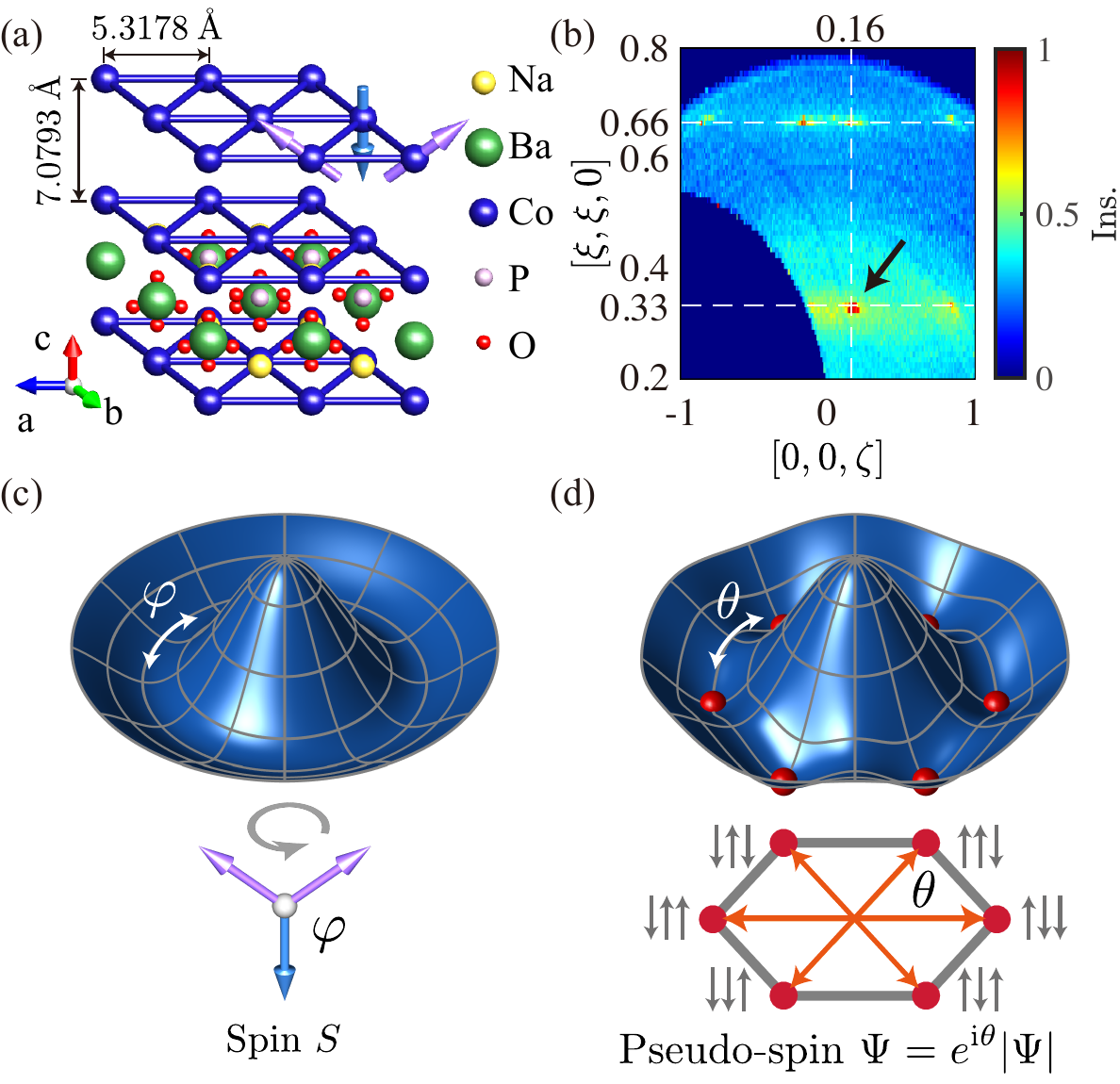}
\caption{(a) Layered triangular-lattice structure of the compound \nbcp {(space group 
P$\bar{3}$m1) with the lattice constants and Y-shape magnetic structure of the spin 
supersolid specified.} (b) The elastic neutron scattering results are obtained at 55 mK, 
from which the lattice constants $a = b = 5.3178~{\rm \AA}$ and $c = 7.0793~{\rm \AA}$ 
are determined. The {arrow} indicates the strongest magnetic peak with the propagation 
vector of $k$ = (1/3, 1/3, 0.16).  
(c) The angle $\varphi$ represents the U(1) phase of spin superfluidity, 
and (d) $\theta$ is the U(1) phase of the {pseudo-spin $\Psi$ of spin solid order} 
(see definition in the main text), which are related to the gapless Goldstone 
and gapped pseudo-Goldstone modes, respectively. 
}
\label{Fig1}
\end{figure}
% ======================== %

Recently, an easy-axis Co-based triangular-lattice compound \nbcp~(NBCP) 
has raised great attention~\cite{Zhong2019,LiN2020,Lee2021,Wellm2021,
Gao2022Super,Wu2022PNAS,Huang2022thermal,Xiang2024Nature}. 
The Co$^{2+}$ ions form stacked triangular lattices [c.f., Fig.~\ref{Fig1}(a)], 
which carry effective spin $S=1/2$ under the effects of spin-orbit coupling 
and crystal electric field. The spin-spin couplings are highly two-dimensional, 
i.e., the intra-layer spin exchanges are dominating over those between the 
layers~\cite{Zhong2019,Wellm2021,Gao2022Super,Wu2022PNAS}. The 
highly frustrated compound NBCP was initially proposed to host a QSL, 
where magnetic ordering was absent down to 300~mK~\cite{Zhong2019,
Lee2021}. A later specific heat measurement finds an anomaly at about 
$150$~mK~\cite{LiN2020,Huang2022thermal}, which instead suggests the 
formation of certain spin order at low temperature. Based on the tensor-network 
calculations of an easy-axis TLAF model, it has been proposed theoretically
\cite{Gao2022Super} and observed experimentally in NBCP the long-sought 
spin supersolid~\cite{Xiang2024Nature}, i.e., a quantum superposition of 
the spin solid and superfluid, spontaneously breaking both the lattice translation 
and the U(1) rotation symmetries~\cite{Wessel2005,Melko2005,Heidarian2005,
Prokofev2005,Heidarian2010,WangF2009,Jiang2009}. The thermodynamics
perspective of spin supersolid has been investigated, where a giant magnetocaloric 
effect (MCE) has been discovered~\cite{Xiang2024Nature}. 

In spin supersolid, there exists a spin current that flows without dissipation. 
This phenomenon is supposed to be closely related to the elementary excitations of the system.
Nevertheless, the dynamical properties of spin supersolids, particularly the low-energy 
excitations, remain largely unexplored in both experiments and theories. In helium-4 
quantum superfluid, the distinctive phonon-roton excitations~\cite{Landau1941Theory,
Landau1947Theory,Feynman1954,Feynman1956,Palevsky1957Roton,Yarnell1959Excitation,
Fukushima1989,Glyde2018Review,Godfrin2021Roton}, first hypothesized by Landau
\cite{Landau1941Theory,Landau1947Theory} and further substantiated by Feynman
\cite{Feynman1954,Feynman1956}, play an essential role in determining its thermodynamic 
and hydrodynamic properties~\cite{Kramers1952,Bendt1959}. As the spin supersolid 
contains both solid and superfluid ``components'', a compelling question thus arises: 
What are the excitation characteristics of spin supersolid? Furthermore, does a magnetic 
analog to phonon-roton excitations exist? These questions are crucial to address, as spin 
excitations can provide significant insights into the low-temperature thermodynamics, 
e.g., the specific heat, entropy, and thus MCE of spin supersolid.

Here we perform tensor-network calculations
on the easy-axis TLAF model for NBCP, complemented by spin-wave calculations, 
and conduct INS measurements on single-crystal samples down to 
55~mK. Our tensor-network results reveal distinctive dynamical features, including 
the gapless Goldstone and gapped pseudo-Goldstone modes at the ${\rm K}$ 
point of the Brillouin zone (BZ), together with the gapped roton modes above the 
${\rm M}$ point with a pronounced downward renormalization. Among these
characteristics, the gapless Goldstone mode is a crucial indicator 
of the spin supersolidity in the easy-axis, Ising-like antiferromagnet, which can be  
observed in the present INS measurements. Synergy between experiments and 
simulations yields a comprehensive understanding of the dynamical properties of the
spin supersolid phase. Overall, the peculiar double magnon-roton dispersions render strong 
spin fluctuations and large magnetic entropies down to very low temperature, well 
explaining the giant MCE observed in NBCP~\cite{Xiang2024Nature}.

%%%%%%%%%%%%%%
{\textit{Dynamical calculations of the easy-axis TLAF model.---}}
%%%%%%%%%%%%%%
An easy-axis TLAF model has been {proposed to accurately describe} the 
compound NBCP~\cite{Gao2022Super,Xiang2024Nature}. The model Hamiltonian 
reads $H = \sum_{\langle i,j\rangle} J_{xy} (S_i^xS_j^x+S_i^yS_j^y) + J_z S_i^zS_j^z$, 
where the nearest-neighbor interactions between $i, j$ are $J_{xy} = 0.88~$K 
and $J_{z}=1.48~$K. {The model parameters were determined by fitting the 
measured specific heat and magnetic susceptibility data~\cite{Gao2022Super}, 
and also obtained independently by analyzing the INS data in the polarized 
phase~\cite{Wu2022PNAS}. Using the same set of parameters, the isothermal 
magnetization curves measured at 22~mK under out-of-plane fields can be 
accurately reproduced, where the calculated critical fields align excellently with 
experiments~\cite{Lee2021,Gao2022Super,Xiang2024Nature}. Moreover, the 
simulated isentropic lines show a remarkable agreement with adiabatic demagnetization 
cooling measurements even below $100$~mK~\cite{Xiang2024Nature}. This strongly 
supports the validity and accuracy of the model used in the current study of {NBCP}.} 
The Y-shape magnetic structure is shown in Fig.~\ref{Fig1}(a), 
where the out-of-plane components spontaneously break the lattice translation symmetry, 
while the in-plane components break the U(1) rotation symmetry. Through the 
rigorous mapping between the spin-1/2 and hardcore bosons, the Y state
constitutes a quantum magnetic analog of a supersolid state~\cite{Gao2022Super,
Wessel2005,Melko2005,Heidarian2005,Prokofev2005,Heidarian2010,WangF2009,Jiang2009}, i.e., a spin supersolid.

% ============ Fig2 ============ %
\begin{figure*}[!htbp]
\includegraphics[angle=0,width=0.9\linewidth]{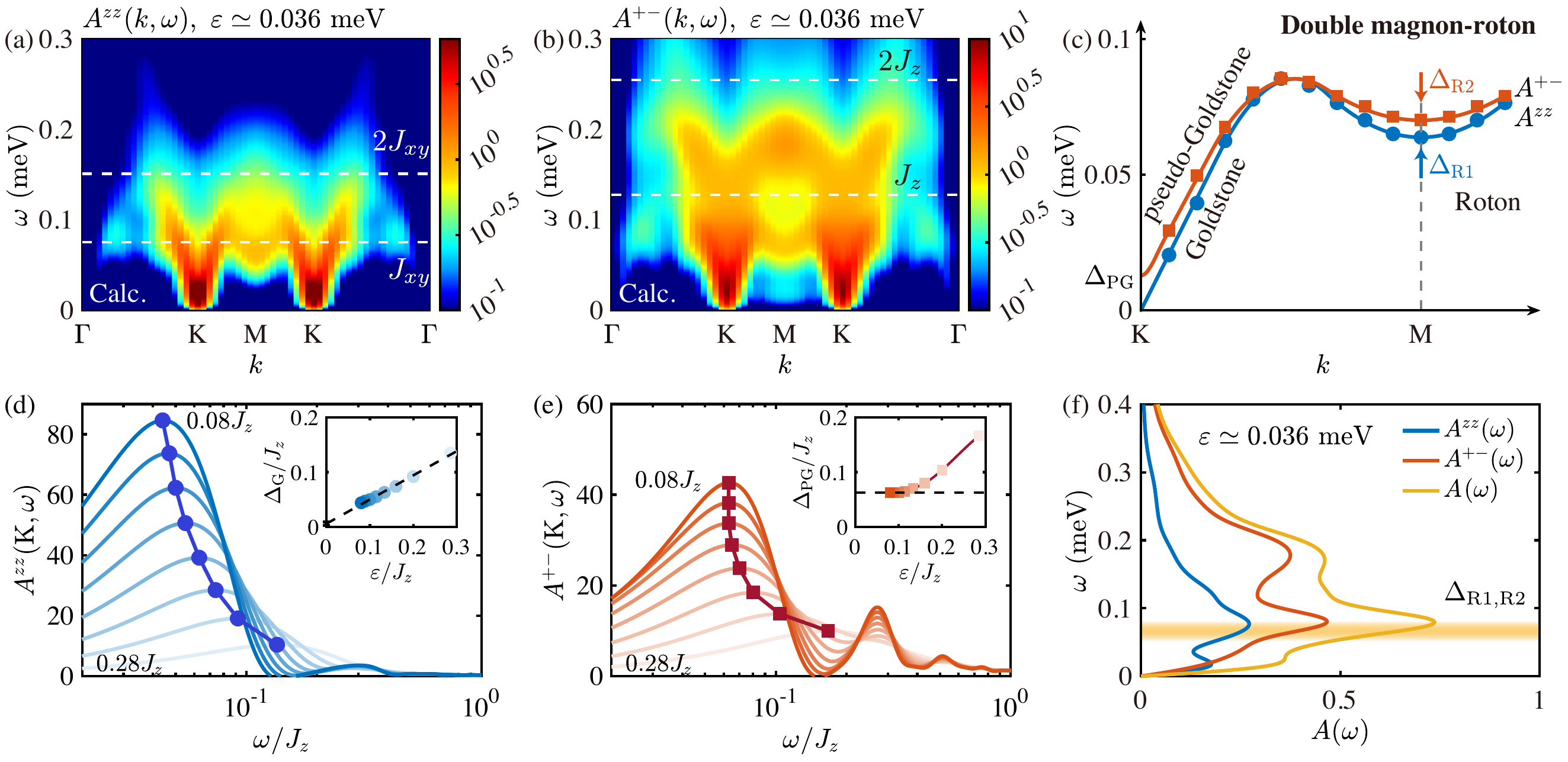}
\caption{The calculated spin-resolved spectral functions (a) $A^{zz}(k, \omega)$ and 
(b) $A^{+-}(k, \omega)$ under zero field, with a high energy resolution of $\varepsilon 
\simeq 0.036~{\rm meV}$. (c) illustrates the double magnon-roton excitations determined 
from the calculated spectral functions. The former corresponds to a dispersion with 
gapless Goldstone mode and roton minimum with gap $\Delta_{\rm R1} \simeq 
0.064~{\rm meV}$, while the latter has a pseudo-Goldstone gap $\Delta_{\rm PG}$ 
and roton gap $\Delta_{\rm R2} \simeq 0.07~{\rm meV}$. The dispersions labeled by 
solid circles and squares are determined from $A^{zz}(k,\omega)$ and 
$A^{+-}(k, \omega)$, respectively. (d) shows the spectral function 
$A^{zz}(k, \omega)$ at $k = {\rm K\equiv(1/3, 1/3)}$ with different energy resolution 
$\varepsilon$ ranging from $0.28 J_z$ (i.e., about 0.036~meV) to $0.08J_z$ 
(about 0.01~meV), and the inset shows that the gap $\Delta_{\rm G}(\varepsilon$) 
extrapolates to zero as $\varepsilon \rightarrow 0$. 
(e) depicts $A^{+-}(k = {\rm K}, \omega)$, from which a finite gap $\Delta_{\rm PG}(0) 
\simeq 0.063 J_z$ can be estimated, as illustrated in the inset. (f) shows the local 
spectral function {$A^{zz}(\omega)$, $A^{+-}(\omega)$, and $A(\omega)$ as 
the sum of the two. The orange shaded area represent the estimated roton gap 
$\Delta_{\rm R1, R2}$. }}
\label{Fig2}
\end{figure*}
% ============================ %

Below we compute the dynamical spin 
structure factor $S(k, \omega)$ and spectral function $A(k, \omega)$. These two 
quantities can be obtained from the real-time correlation function {$g^{\alpha\beta}(k,t)
= \frac{1}{N} \sum_{i,j} e^{-{{\rm i} k\cdot(r_i-r_j)}} g^{\alpha \beta}_{ij}(t)$, 
where $g^{\alpha \beta}_{ij}(t) = e^{{\rm i} E_{0} t} \bra{\psi_0} 
S^{\alpha}_i e^{-{\rm i} H t} S^{\beta}_j \ket{\psi_0} $, }
$\ket{\psi_0}$ ($E_0$) 
is the ground-state wavefunction (energy), and $N$ is the total {number of sites}. 
The real-time evolution of the {quenched} ground state 
$\ket{\psi_0}$ is computed via {the matrix product state (MPS) approach} 
based on time-dependent variational principle~\cite{White1992,TDVP2011,TDVP2016, Supplementary}. 
Given the correlation function $g^{\alpha\beta}(k,t)$, the spin-resolved dynamical
structure factor and spectral function can be computed as $S^{\alpha\beta}(k,\omega) 
= \int_0^{t_{\rm max}} {\rm Re} [g^{\alpha\beta}(k,t) e^{\rm{i}\omega t}] \, 
W({t}/{t_{\rm max}}) \, dt$, and $A^{\alpha\beta}(k,\omega) = -\int_0^{t_{\rm max}} 
{\rm Im} [g^{\alpha\beta}(k,t)] \sin(\omega t) \,W({t}/{t_{\rm max}}) \, dt$, respectively. 
$t_{\rm max}$ controls the energy resolution $\varepsilon \simeq 8/t_{\rm max}$ 
when the Parzen window function $W({t}/{t_{\rm max}})$ is used~\cite{Supplementary}. 
In practice, the dynamical calculations are performed on {YC$W\times L$ 
cylinders with width $W$ and length $L$, with evolution 
time up to $t_{\max} = 100/J_z$. We adopt the MPS-based approach and
capture the entire excitation spectrum ranging from low- to high-energy regimes
(see Supplementary Materials~\cite{Supplementary}). {Symmetries are implemented 
using the highly efficient tensor-network package QSpace~\cite{QSpace,weichselbaum2024QSpace}.

In Figs.~\ref{Fig2}(a,b) we show the spin-resolved spectral results of 
$A^{zz}(k,\omega)$ and $A^{+-}(k,\omega)$, which exhibit rather distinct 
behaviors. The spectral functions do not include the elastic-scattering 
peaks and allow us to concentrate on the low-energy fluctuations. From 
Fig.~\ref{Fig2}(a) we find $A^{zz}(k,\omega)$ intensities are significant 
only for $\omega \lesssim 0.15$~meV, i.e., below about $2 J_{xy}$. On 
the other hand, $A^{+-}(k,\omega)$ can extend to higher energies of 
about $0.25~{\rm meV} \sim 2J_z$. In Figs.~\ref{Fig2}(a,b), we find 
clear magnetic excitation dispersions for $\omega \lesssim J_{xy, z}$,
dub as magnon-roton dispersions, above which there are signatures 
for excitation continuum (see Supplementary Materials~\cite{Supplementary}). 

\textit{Goldstone, pseudo-Goldstone, and roton modes.---}
To examine the low-energy excitations, we gradually improve the energy resolution 
to about $0.08J_z$, and find $A^{zz}(k, \omega)$ becomes more coherent as 
$\varepsilon$ decreases, as shown in Fig.~\ref{Fig2}(d). Upon convolution with 
window functions, the peak location $\Delta_{\rm G}$ of $A^{zz}(k, \omega)$ has 
been shifted to higher frequencies. As shown in the inset of Fig.~\ref{Fig2}(d), 
we find $\Delta_{\rm G}$ becomes lowered as $\varepsilon$ decreases, and it 
extrapolates eventually to approximately zero energy in the $\varepsilon=0$ limit. 
It clearly indicates the existence of in-plane gapless Goldstone modes as illustrated 
in Fig.~\ref{Fig1}(c). 

On the other hand, in Fig.~\ref{Fig2}(b) we reveal the low-energy out-of-plane 
excitations by computing the spectral function $A^{+-}({k}, \omega)$. 
As shown in Fig.~{\ref{Fig2}}(e), a small but nonzero gap $\Delta_{\rm PG} 
\simeq 0.063J_z$ exists, which is related to the pseudo-Goldstone mode 
as illustrated by the ``{distorted} Mexican hat'' energy landscape with 
six-fold degenerate ground states in Fig.~\ref{Fig1}(d). 
To see that, we can introduce the complex order parameter {$\Psi = 
e^{\rm i \theta} |\Psi| \equiv \frac{1}{N} (\sum_{i\in A} \langle S^z_{i} \rangle
+ \sum_{j\in B} \langle S^z_{j}\rangle e^{{\rm i} \pi 2/3}+ \sum_{k \in C} \langle 
S^z_{k} \rangle e^{{\rm i} \pi 4/3})$}, where $A,B,C$ label the three sublattices 
and the U(1) phase $\theta$ reveals the ``hidden'' XY degree of freedom. 
In Fig.~\ref{Fig1}(d), the spin configurations like $\uparrow \uparrow \downarrow$ 
and $\uparrow \downarrow \uparrow$, etc., correspond to the six-fold degenerate 
ground state with $\theta = n \pi/3$ ($0 \leq n \leq 5$). The transverse excitation 
modes are now gapped near the six minima, and this small pseudo-Goldstone gap 
is generated by quantum fluctuations via the order-by-quantum-disorder mechanism
\cite{Rau2018}. {Similar to the triangular-lattice Ising model under transverse 
field~\cite{Isakov2003XY,Li2020TMGO,Hu2020TMGO}, the six-fold anisotropy 
in pseudo-spin $\Psi$ [see Fig.~\ref{Fig1}(d)] becomes irrelevant above certain 
temperature, and there emerges a Berezinskii-Kosterlitz-Thouless phase~\cite{Berezinskii1971,
Kosterlitz1972} with emergent U(1) symmetry.}

Despite intensive theoretical~\cite{Chubukov1994,Zheng2006PRL,Zheng2006,
Verresen2019Avoided,Chen2019} and experimental investigations
\cite{Ma2016Dynam,Ito2017Dynam,Macdougal2020} on the roton-like minima in spin 
excitations of (nearly) isotropic Heisenberg TLAF systems, significantly less is 
understood about the easy-axis case that exhibits a spin supersolid phase. {We 
summarize the results} in Fig.~\ref{Fig2}(c), where two branches of excitations are seen. 
The in-plane excitations contain a linear dispersion and soft quadratic excitation near the 
${\rm M} \equiv (1/2, 1/2)$ point, serving as a magnetic analog of Landau's phonon-roton 
excitations in superfluid helium~\cite{Landau1941Theory,Landau1947Theory}. Remarkably, 
in Fig.~\ref{Fig2}(c) there is a second magnon-roton dispersion {that can be associated 
with the fluctuations of spin-solid order characterized by $\Psi$.} 

%%%%%%%%%%%%%%
{\textit{Samples and neutron scattering measurement.---}}
Single-crystal samples of \nbcp~were grown using the flux method as reported 
in Ref.~\onlinecite{Xiang2024Nature}. The INS experiments were performed on 
the cold-neutron time-of-flight spectrometer PELICAN {at ANSTO~\cite{Pelican}}. 
A total number of 28 pieces of NBCP single crystals with a total mass of about 
3~g were mounted on the sample holder. 
The base temperature of 55~mK was achieved using a dilution insert inside a 
7 T vertical cryomagnet. The crystals were co-aligned with their [1, $-$1, 0] 
direction lying vertically, so that the [$H$, $H$, $L$] scattering plane can be 
mapped out by rotating the sample horizontally and an in-plane field can be 
applied along the [1, $-$1, 0] direction.
The instrument was configured with an incident neutron wavelength of $5.96$~\AA, 
providing an incident energy of $2.3$~meV with a high energy resolution of about 
$0.066$~meV at the elastic line. A standard vanadium sample was measured 
for detector normalization and determination of the energy resolution function.
The data reductions 
were performed using the software HORACE~\cite{Horace}

% =============== Fig3 ============== %
\begin{figure}[]
\includegraphics[angle=0,width=1\linewidth]{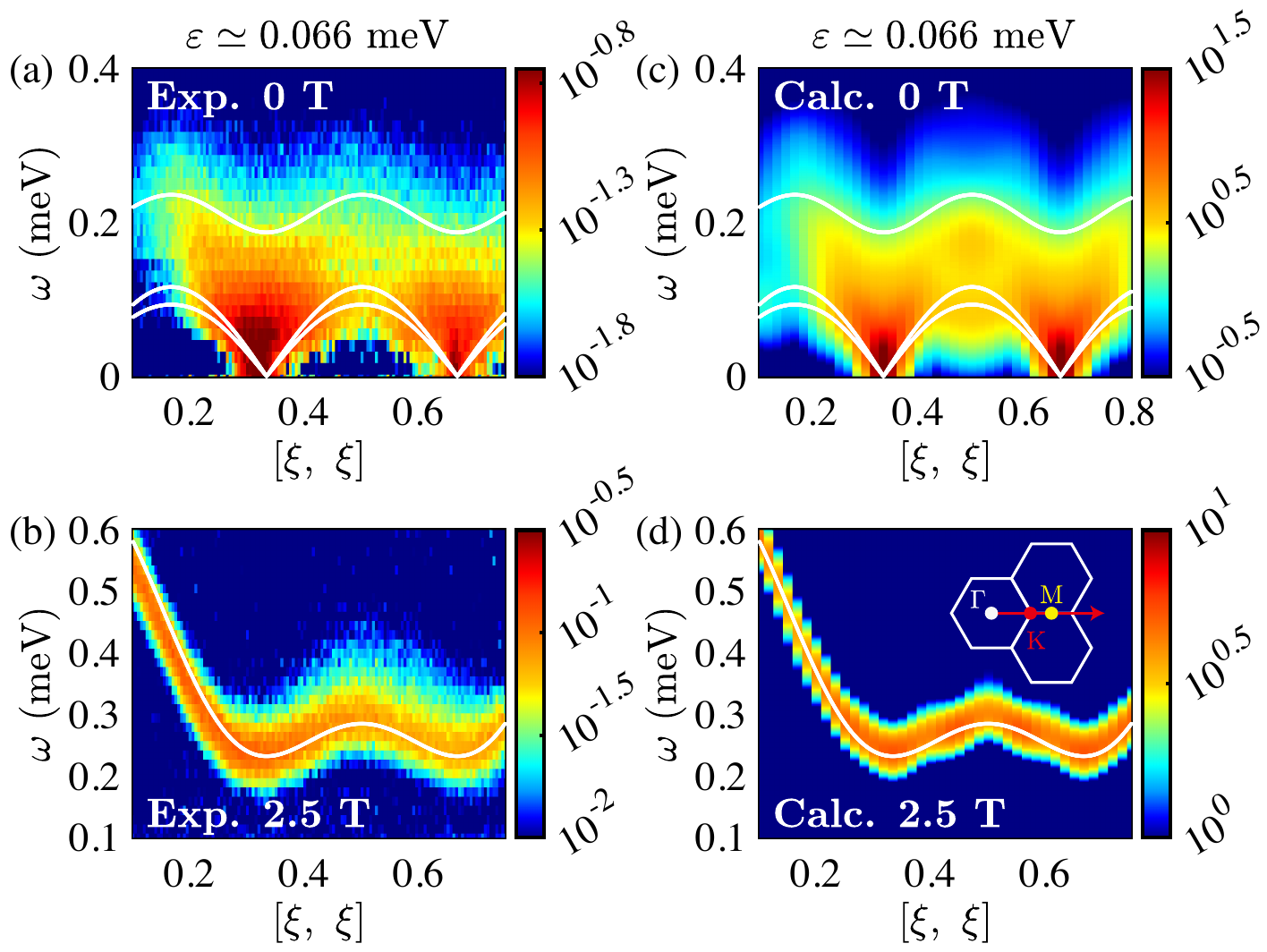}
\caption{INS data under (a) zero and (b) 2.5~T in-plane fields, which are measured 
along the path $[\xi,\xi]$ (red arrow) shown in the inset of (d).
Scatterings are integrated along the $[-\eta, \eta, 0]$ direction perpendicular to the 
horizontal scattering plane for $\eta \in {[-0.1,0.1]}$ and along the [0, 0, $\zeta$] 
direction for $\zeta \in {[-1,1]}$. The dataset below 0.15 meV collected at $2.5$~T 
is used as background subtraction for the 0~T case \cite{Supplementary}. 
(c) and (d) show the calculated dynamical spin 
structure factor $S(k,\omega) = \sum_{\alpha=\{x,y,z\}} S^{\alpha\alpha}(k,\omega)$ 
under zero and 2.5~T fields, respectively. {The results in (d) are obtained under
an in-plane field, i.e., $g_{\rm ab} \mu_{\rm B} B \sum_i S_i^x$, with the Land\'e factor 
$g_{\rm ab} \simeq 4.24$ and $\mu_B$ the Bohr magneton.}  
The LSW results are shown with solid white lines.}
\label{Fig3}
\end{figure}
% =============================== %

%%%%%%%%%%%%%%
{\textit{Spectroscopic evidence for spin supersolid.---}}
%%%%%%%%%%%%%%
In Fig.~\ref{Fig1}(b) we present the elastic scattering results (within 
$\omega=0\pm0.05$ meV)  of the {raw} INS measurements, which clearly 
reveal magnetic Bragg peaks at ${\rm K} \equiv (1/3, 1/3)$ with an 
incommensurate out-of-plane propagation vector {$k_z$} = 0.16.
The latter may be ascribed to the high sensitivity of the spin supersolid 
state to weak interlayer couplings {due to its gapless nature}~\cite{Xiang2024Nature}. 
Further analysis shows that the magnetic Bragg peaks are contributed 
mainly from the out-of-plane moments~\cite{Supplementary}, 
consistent with prior measurements and analysis~\cite{Wu2022PNAS,
Xiang2024Nature}. Besides, the low-energy spin fluctuations exhibit a 
rod-like shape, {as better visualized in Supplementary Materials~\cite{Supplementary},} 
indicating very good two dimensionality of NBCP.

In Figs.~\ref{Fig3}(a,b) we present the low-energy magnetic excitations 
observed at $55~{\rm mK}$ under zero and $2.5~{\rm T}$ in-plane fields, 
and compare them to the model calculations. In particular, gapless Goldstone 
modes are evidenced in Fig.~\ref{Fig3}(a) under zero field, where linear 
spin-wave (LSW) dispersions emanating from the ordering vector ${\rm K}$ are 
also plotted for comparison. The tensor-network calculations with realistic 
model parameters and similar energy resolution well agree with the experimental 
measurements. These results support the existence of gapless Goldstone 
mode (see the energy cut across the ${\rm K}$ point in the Supplementary 
Materials~\cite{Supplementary}), which is 
critical for identifying spin supersolidity in the easy-axis antiferromagnet NBCP.} 
On the other hand, $B=2.5~{\rm T}$ exceeds the in-plane 
{saturation field $B_s \simeq 1.51$~T}~\cite{LiN2020,Gao2022Super}, 
and a clear magnon dispersion in the nearly polarized phase can be observed. 
{As shown in Figs.~\ref{Fig3}(b,d)}, the theoretical results, {including 
both tensor-network and LSW calculations,} demonstrate an excellent 
match with experiment, confirming once again the precision of the easy-axis 
TLAF model for NBCP~\cite{Gao2022Super}.

Moreover, Fig.~\ref{Fig3}(a) reveals the presence of an extra intensity that overlays 
the spin-wave dispersion, notably concentrated around the $\mathrm{M} \equiv (1/2, 1/2)$ 
point of the BZ, which is clearer in the energy cut shown in the Supplementary 
Materials~\cite{Supplementary}. Such a downward renormalization can be discerned 
in experimental data in Fig.~\ref{Fig3}(a) and more clearly in the simulated results in 
Fig.~\ref{Fig3}(c). There are significant spectral weights at energies lower than the 
corresponding LSW excitations. This downward renormalization can be regarded as 
signature of roton modes --- magnetic analog of rotons in superfluid helium
\cite{Landau1941Theory,Landau1947Theory,Feynman1954,Feynman1956,Glyde2018Review}.
The existence of roton mode is rather clear from model calculations [see Figs.~\ref{Fig2}(a,b)], 
while its experimental observation is quite challenging for NBCP due to its very 
low energy scale.

% ========== Fig4 ========= %
\begin{figure}[]
\includegraphics[angle=0,width=1\linewidth]{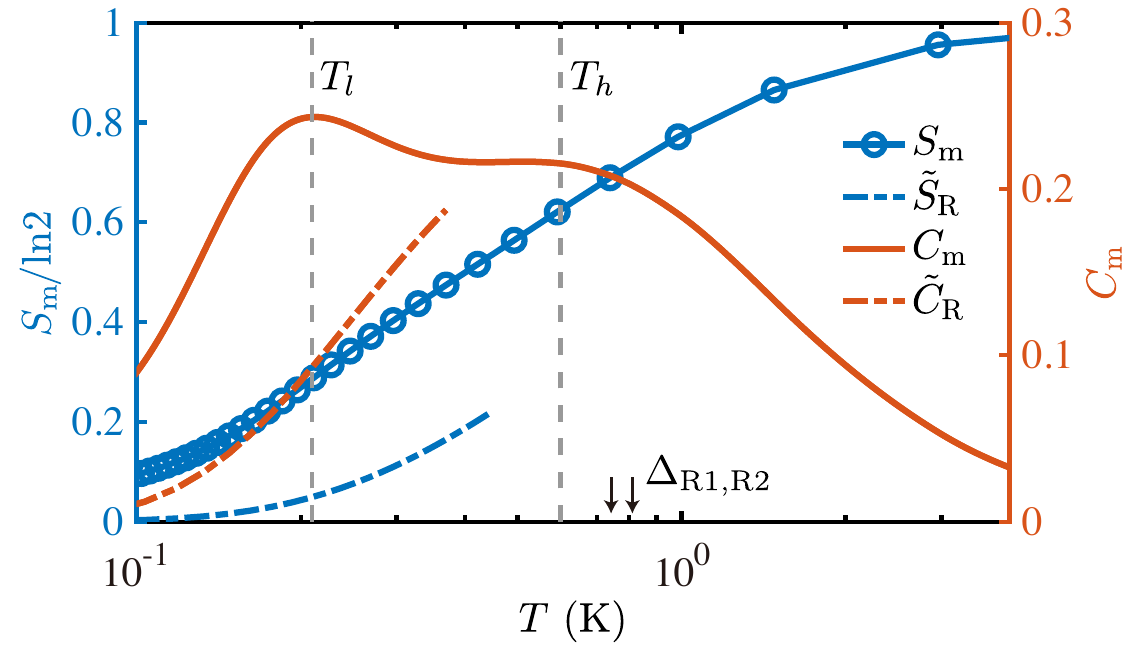}
\caption{The specific heat $C_{\rm m}$ and total magnetic entropy 
$S_{\rm m}$ are shown, 
which are computed on a YC6$\times$15 cylinder with tanTRG
\cite{tanTRG2023} method with $D=5000$ bond states retained. 
The higher ($T_h\simeq 0.6~{\rm K}$) and the lower ($T_l \simeq 
0.2~{\rm K}$) temperature scales are determined from the two peaks 
in the specific heat. The roton entropy contributions $\tilde{S}_{\rm R}$ 
and roton specific heat $\tilde{C}_R$ are estimated based on the 
spectral density of roton modes in Fig.~\ref{Fig2}(f). Roton gaps 
$\Delta_{\rm R1}$ and $\Delta_{\rm R2}$ are also indicated in the plot.
}
\label{Fig4}
\end{figure}
%% ======================== %

%%%%%%%%%%%%%%
{\textit{Thermodynamics of magnon-roton excitations.---}}
{For superfluid helium with phonon-roton excitations, the rotons can be activated at a 
temperature much lower than its gap and contribute significantly to the low-temperature 
thermodynamics, due to their anomalously large density of states~\cite{Kramers1952}. 
Here for the spin supersolid with double magnon-roton excitations, we show in 
Fig.~\ref{Fig2}(f) the normalized local spectral density $A(\omega) =  \frac{1}{\mathcal{N}}
(A^{zz}(\omega) + A^{+-}(\omega))$, where $A^{\alpha\beta}(\omega)$ is derived 
from the local correlation function $g^{\alpha\beta}_{ii}(t)$ and $\mathcal{N}$ is the 
normalization factor~\cite{Supplementary}. A prominent peak 
is observed in $A(\omega)$ near the roton gap $\omega \sim \Delta_{\rm R1, R2}$, 
which can strongly influence the low-temperature properties.}

In Fig.~\ref{Fig4}, we show the calculated specific heat $C_{\rm m}$ and magnetic entropy $S_{\rm m}$ 
from thermal tensor networks. By taking $A(\omega)$ as the effective density of states 
{(with $\omega\in[0.04, 0.1]~{\rm meV}$)}~\cite{Supplementary}, 
we provide an estimation, albeit somewhat rudimentary, of roton contributions in $\tilde{C}_{\text{R}}$ 
and $\tilde{S}_{\text{R}}$ at low temperature. From the results, one can see that the roton modes 
contribute anomalously large entropy even below $T_l \simeq 0.2$~K, despite the existence roton 
gap $\Delta_{\rm R1, R2} \approx 0.78$~K, reminiscent of the thermodynamics of superfluid helium. 
Therefore, the double magnon-roton excitations significantly increase the low-temperature spin fluctuations 
and naturally explains the giant MCE observed in NBCP~\cite{Xiang2024Nature,Zhang2024PRM}.

%%%%%%%%%%%%%%
{\textit{Discussion and outlook.---}}
{In our comprehensive study of spin dynamics in the easy-axis TLAF system, we obtain the double 
magnon-roton excitations through tensor-network simulations, and observe the proposed Goldstone 
modes as well as signatures of rotons in the INS measurements. Our results not only provide 
spectroscopic evidence for spin supersolidity in NBCP, but also shed light on other supersolid 
candidates~\cite{Sengupta2007field,Sengupta2007chain,Mila2008SS,Wierschem2013,Shi2022,
Wang2023SS}. In particular, recently people find experimental evidence for spin supersolid 
in the compound K$_2$Co(SeO$_3$)$_2$,} despite a different extent of easy-axis anisotropy
\cite{RCSO2020,zhu2024continuum,chen2024phase}. Owing to the significantly larger spin 
exchange in this cobaltate, {$J_z \approx 3$~meV~\cite{zhu2024continuum,chen2024phase},} 
we anticipate that observing the predicted dynamical features, including the roton modes and 
pseudo-Goldstone gap, may require less stringent experimental conditions compared to \nbcp. 

{Very recently, there is a notable interest in the excitation continuum within the supersolid 
phase~\cite{chen2024continuum,sheng2024continuum,zhu2024continuum,chen2024phase,
xu2024KCSO,chi2024dynamical,ulaga2024KCSO}, which is also present in the relatively 
high-energy regime of our calculations, and its origin warrants further exploration 
(see Supplementary Materials~\cite{Supplementary}). Additionally, theoretical proposals 
have been raised for realizing supersolid states with Rydberg atoms~\cite{liu2024Rydberg,
homeier2024Rydberg}, paving the way for exploring their fascinating quantum dynamics through 
quantum simulations.}

\begin{acknowledgements}
\textit{Acknowledgments.---}
W.L., W.J., and Y.G. express their gratitude to Tao Shi, Changle Liu, Shang Gao, 
Bing Li, and Hai-Jun Liao for stimulating discussions. This work was supported 
by the National Natural Science Foundation of China (Grant Nos.~12222412, 
12047503, 12074023, {11834014}), National Key Projects for Research and Development 
of China (Grant Nos.~{2018YFA0305800}, 2021YFA1400400, 2023YFA1406003), and the CAS 
Project for Young Scientists in Basic Research (Grant No.~YSBR-057). 
We thank the HPC-ITP for the technical support and generous allocation 
of CPU time. The Australian Center for Neutron Scattering is gratefully 
acknowledged for providing neutron beam time through Proposal No. P17086.
\end{acknowledgements}
\bibliography{../Draft/MCERef.bib}

%%% ######################################################################### %
%%% ######################################################################### %
%%% ########################## SUPPLEMENTARY ################################ %
%%% ######################################################################### %
%%% ######################################################################### %
\newpage
\clearpage
\onecolumngrid
\mbox{}
\begin{center}
\textbf{Supplementary Material for}
\\ {Double Magnon-Roton Excitations in the Triangular-Lattice Spin Supersolid}\\
~\\
Yuan Gao, et al.
\end{center}
\date{\today}
\setcounter{section}{0}
\setcounter{figure}{0}
\setcounter{equation}{0}
\renewcommand{\theequation}{S\arabic{equation}}
\renewcommand{\thefigure}{S\arabic{figure}}
\setcounter{secnumdepth}{3}

\onecolumngrid

\section{Data analysis of the neutron scattering measurements}

The low-energy range with $\omega\in$  [-0.05, 0.05]~meV of the INS data in zero 
field was integrated and treated as the elastic scattering results. As shown in 
Fig.~\ref{FigS_ElasticFitting}(a), the coexistence of bright magnetic Bragg 
peaks with the propagation vector of $k$ = (1/3, 1/3, 0.16) and diffusive rod-like 
scatterings are observed. The latter is along the out-of-plane direction and 
suggest a quasi two-dimensional nature of NBCP.

The integrated intensities of the magnetic Bragg peaks on top of the diffuse scatterings 
were extracted for further analysis. As shown in Fig.~\ref{FigS_ElasticFitting}(b), the 
intensities of the four non-equivalent reflections agree with an up-up-down (UUD) 
configuration of the Co$^{2+}$ moments along the $c$-axis described by the irreducible 
representation $\Gamma_1$. The moment sizes on the $z = 0$ layer are estimated to 
be 0.606(27), $-$0.303(13) and $-$0.303(13) $\mu\rm_B$, for the ``down'', ``up'', and 
``up'' spins in three sublattices, respectively. The results are well consistent with our 
previous neutron diffraction results on NBCP also under zero field~\cite{Xiang2024Nature} 
and supports the presence of out-of-plane spin solidity in the compound.

% ========== FigS1 ========= %
\begin{figure*}[!htbp]
\includegraphics[angle=0,width=0.55\linewidth]{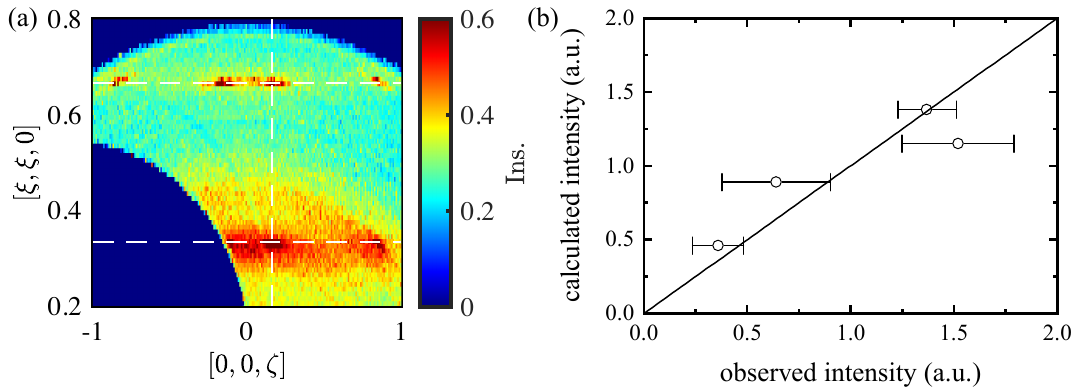}
\caption{(a) Contour plot of the elastic part of INS results with $\omega\in$ 
[$-$0.05, 0.05]~meV. The data is the same with those shown in Fig.~1(a)
in the main text, but with a different color bar to emphasize the rod-like 
diffuse scatterings. They arise from the almost dispersionless spin 
excitations at very low energy along the out-of-plane direction. 
(b) The comparison between the observed and calculated (integrated) 
intensities of four non-equivalent magnetic reflections for $B = 0$~T, 
adopting the UUD spin configuration of the Co$^{2+}$ moments along the 
$c$-axis described by the irreducible representation $\Gamma_1$. }
\label{FigS_ElasticFitting}
\end{figure*}
% ======================== %

By integrating the scatterings along the out-of-plane 
[0, 0, $\zeta$] direction, it is found that the spin excitations emanating and 
away from the ordering vector look quite similar, as shown in { 
Figs.~\ref{FigS_Diffomega}(a,d) for $\zeta\in$  [$-$0.5, 0.5], 
Figs.~\ref{FigS_Diffomega}(b,e) for $\zeta\in$ [$-$0.7, 0.7], and 
Figs.~\ref{FigS_Diffomega}(c,f) for $\zeta\in$  [$-$1, 1],} 
also indicating a very good two dimensionality of NBCP. 

% ========== FigS3 ========= %
\begin{figure*}[!htbp]
\includegraphics[angle=0,width=0.9\linewidth]{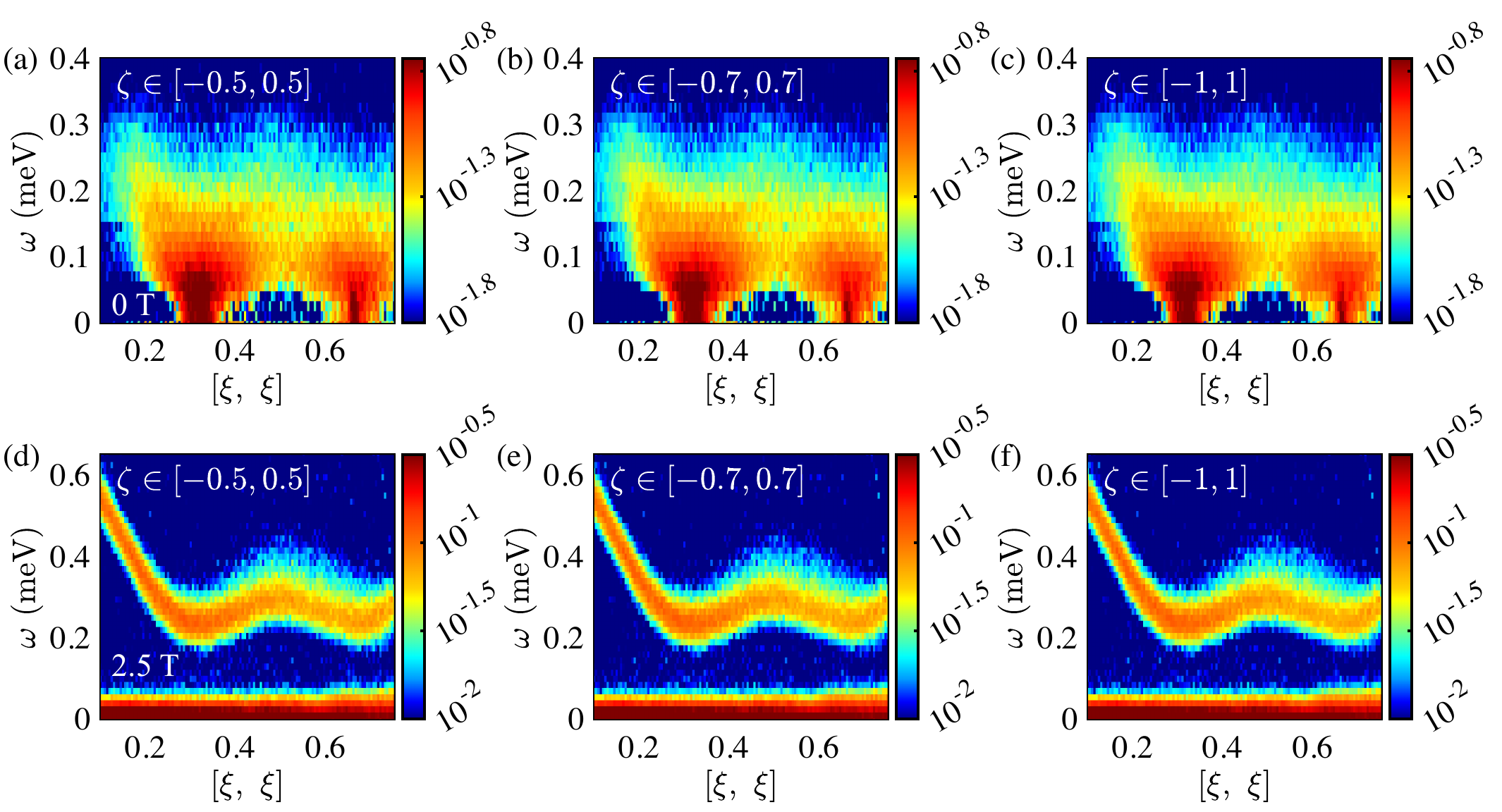}
\caption{The INS data cut along the high-symmetry $[\xi,\xi]$ direction 
measured in (a-c) zero field {and (d-f) 2.5 T in-plane field}. Scatterings 
are integrated along the [$-\eta$, $\eta$, 0] direction for $\eta\in$  {[$-$0.1, 0.1]}, 
and along the out-of-plane [0, 0, $\zeta$] direction for {(a,d) $\zeta\in$  [$-$0.5, 0.5]
, (b,e) $\zeta\in$  [$-0.7$, 0.7], and (c,f) $\zeta\in$  [$-1$, 1] respectively.} 
The close agreement between the two plots indicates that the compound 
exhibits very good two-dimensionality.}
\label{FigS_Diffomega}
\end{figure*}
% ======================== %

In Fig.~\ref{FigS_BK} we show the raw INS data collected at 55 mK under zero 
[Fig.~\ref{FigS_BK}(a)] and 2.5~T in-plane field [Fig.~\ref{FigS_BK}(b)]. As shown 
in Fig.~\ref{FigS_BK}(b), the spin excitations under an in-plane field of $B=2.5~{\rm T}$ 
are clearly gapped. Therefore, we utilize the low-energy part ($\omega\in$  [0, 0.15]~meV) 
of the 2.5 T data as the background to be subtracted from the 0~T data, and obtain the 
result shown in Fig.~3(a) of the main text. 

% ========== FigS3 ========= %
\begin{figure}[htbp]
\includegraphics[angle=0,width=0.55\linewidth]{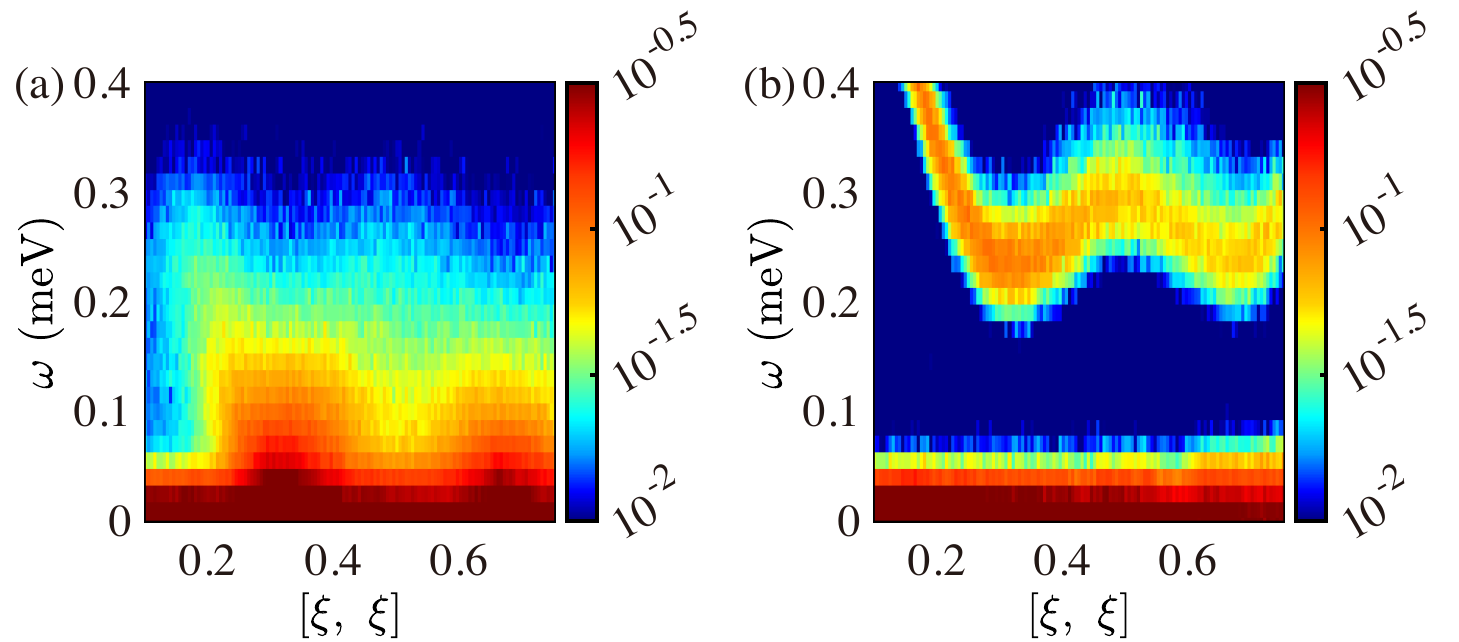}
\caption{Raw INS data measured at 55 mK under (a) $B=0$ and 
(c) $B=2.5~{\rm T}$ in-plane field.}
\label{FigS_BK}
\end{figure}
% ======================== %

Additionally we show the energy cuts of the dynamical spin structure factors 
across the ${\rm K}$ and ${\rm M}$ points in Figs.~\ref{FigEM_cut}(a) and (b) respectively. 
The experimental data and model calculations exhibit good agreement at 
low energy, although quantitative discrepancies emerge at higher energy. 
The INS experiments and tensor-network calculations consistently demonstrate
the gapless and gapped excitations at the ${\rm K}$ and ${\rm M}$ points, 
respectively. Notably, there is a great agreement in the roton gap obtained 
in experiments and tensor networks, which is smaller than the LSW results, 
indicating thus a clear downward renormalization of the excitations near ${\rm M}$ point.
At the ${\rm K}$ point, the calculated dynamical structure factor diverges
more rapidly than the experimental data. This may partly be ascribed to the 
fact that the calculations are done at zero temperature while the INS 
measurements are performed at $T= 55$~mK. Additionally, imperfections 
in the samples could also contribute to this difference.

% ========== FigS4 ========= %
\begin{figure}[]
\includegraphics[angle=0,width=0.55\linewidth]{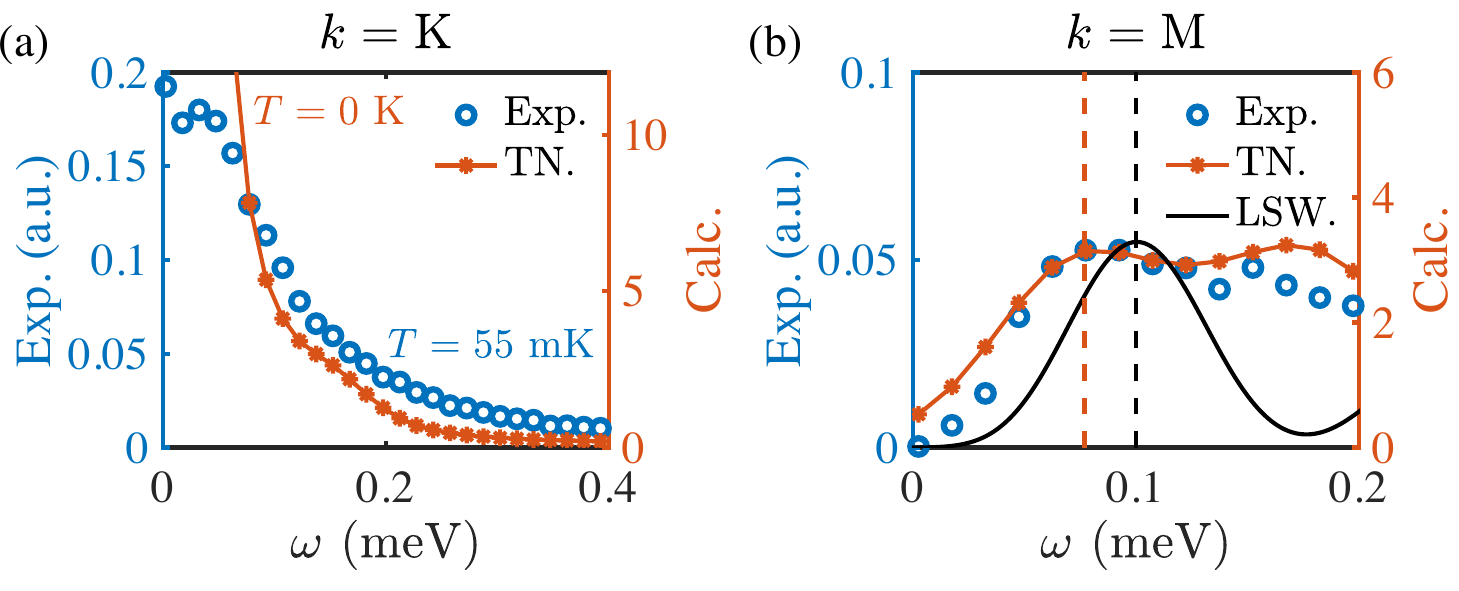}
\caption{Energy cuts of the dynamical spin structure $S(k,\omega)$ at (a) 
${\rm K}$ and (b) ${\rm M}$ point of the BZ. The experimental data are 
measured at 55 mK and the model calculations are performed at 0 K with 
the tensor-network and LSW methods. The energy resolution used in the calculations 
is $\varepsilon \simeq 0.066$~meV. The red and gray arrows indicate the tensor-network 
and LSW results of the roton gap, respectively. The downward renormalization of 
energy is estimated as $0.19 J_z$. a.u. stands for arbitrary unit in the INS data.}
\label{FigEM_cut}
\end{figure}
% ======================== %

\section{Ground-State Dynamical Calculations}
\subsection{Derivation of dynamical spin structure factor and spectral function}
In this section, we detail the derivation of ground-state dynamical spin structure 
factor $S(k,\omega)$ and spectral function $A(k,\omega)$. We start with the 
real-time correlation function 
\begin{equation}
g^{\alpha\beta}(k,t)\equiv \frac{1}{N} e^{{\rm i} E_{0} t} 
\sum_{i,j} e^{-{{\rm i} k\cdot(r_i-r_j)}} \bra{\psi_0}S^{\alpha}_i e^{-{\rm i} H t} 
S^{\beta}_j \ket{\psi_0} = {e^{{\rm i} E_{0} t}  \bra{\psi_0}S^{\alpha}_{-k} e^{-{\rm i} H t} 
S^{\beta}_{k} \ket{\psi_0}},
\label{EqS:gkt}
\end{equation}
where $\ket{\psi_0}$ is the ground state, $E_0$ is the ground-state energy, 
$N$ is the total site number, and $(S_i^\alpha)^{\dagger} = S_i^{\beta}$. 
Then we take the complex conjugate and arrive at
\begin{equation}
g^{\alpha\beta}(k,t)^* = \frac{1}{N} e^{-{\rm i} E_{0} t} 
\sum_{i,j} e^{{-{\rm i} k\cdot(r_j-r_i)}} \bra{\psi_0}S^{\alpha}_j e^{{\rm i} H t} 
S^{\beta}_i \ket{\psi_0} = g^{\alpha\beta}(k,-t).
\end{equation}
With the real-time correlation function, the dynamical spin structure factor $S(k,\omega)$
can be computed as
\begin{equation}
\begin{split}
S^{\alpha\beta}(k,\omega) &\equiv \int_{-\infty}^{\infty} g^{\alpha\beta}(k,t) e^{{\rm i}\omega t}~dt 
=  \int_{0}^{\infty} g^{\alpha\beta}(k,t) e^{{\rm i}\omega t}~dt + 
\int_{0}^{\infty} g^{\alpha\beta}(k,-t) e^{-{\rm i}\omega t}~dt \\
&= 2 \int_{0}^{\infty} {\rm Re}[g^{\alpha\beta}(k,t) e^{{\rm i} \omega t}]~dt.
\end{split}
\end{equation}
Similar, we can calculate the spectral function $A(k, \omega) \equiv 
-\frac{1}{\pi} {\rm Im} [G^{\rm R}(k,\omega)]$ where $G^{\rm R}(k,\omega)$ 
is the retarded Green's function 
\begin{equation*}
\begin{split}
G^{{\rm R}, \alpha\beta}(k,\omega) \equiv & -\frac{\rm i}{N}\sum_{i,j} e^{{-{\rm i} k\cdot(r_i-r_j)}} 
\int_{0}^{\infty} e^{{\rm i}\omega t} \bra{\psi_0}[S^\alpha_i(t), S^\beta_j(0)]\ket{\psi_0}~dt\\
=& -\frac{\rm i}{N}\sum_{i,j} e^{{-{\rm i} k\cdot(r_i-r_j)}} 
\int_{0}^{\infty} e^{{\rm i}\omega t} (e^{{\rm i}E_0 t} \bra{\psi_0}S^\alpha_i 
e^{-{\rm i}Ht} S^\beta_j\ket{\psi_0} - e^{-{\rm i}E_0 t} \bra{\psi_0}S^\beta_j 
e^{{\rm i}Ht} S^\alpha_i\ket{\psi_0})~dt\\
=& -{\rm i}\int_{0}^{\infty} e^{{\rm i}\omega t} [g^{\alpha\beta}(k,t) - g^{\beta\alpha}(-k, -t)]~dt.
\end{split}
%\label{EqS:Grkw}
\end{equation*}
As the Hamiltonian and ground state are invariant under the spatial reversal 
$r_i \rightarrow -r_i$, we have $g^{\alpha\beta}(k,t) = g^{\alpha\beta}(-k,t)$. 
Besides, with zero magnetic field, the Hamiltonian is invariant under the spin 
flip $S_i^z \rightarrow -S_i^z, S_i^y \rightarrow -S_i^y$, thus we have $g^{+-}(k,t) 
= g^{-+}(k,t)$. Considering $A^{+-}(k,\omega), A^{zz}(k,\omega)$ under zero 
field, we obtain the spectral function as 
\begin{equation}
\begin{split}
A^{\alpha\beta}(k, \omega) \equiv& -\frac{1}{\pi} {\rm Im} 
[G^{{\rm R},\alpha\beta}(k,\omega)]
= -\frac{2}{\pi} {\rm Im}  [\int_{0}^{\infty} [\cos(\omega t) + {\rm i} 
\sin(\omega t)] \, {\rm Im}[g^{\alpha\beta}(k,t)]]\\
=& -\frac{2}{\pi} \int_0^{\infty} {\rm Im}[g^{\alpha\beta}(k,t)] \sin(\omega t)~dt.
\end{split}
\label{EqS:Akw}
\end{equation}

In the calculations, we first obtain $\ket{\psi_0}$ using density matrix renormalization 
group~\cite{White1992} and then perform real-time evolution to simulate {$\ket{{\psi}'(t)}
\equiv e^{-{\rm i} H t} S^{\beta}_k \ket{\psi_0}$ [following the last equality in Eq.~(\ref{EqS:gkt})]} 
with time-dependent variational principle 
(TDVP) approach~\cite{TDVP2011,TDVP2016}. Having acquired the real-time correlation 
function, we proceed to calculate the dynamical spin structure factor and the spectral 
functions, which are convolved with an appropriate window function to account for 
a finite energy resolution, i.e.,
\begin{equation}
\begin{split}
S^{\alpha\beta}(k,\omega) &= \int_0^{t_{\rm max}} {\rm Re} 
[g^{\alpha\beta}(k,t) e^{\rm{i}\omega t}] ~ W(\frac{t}{t_{\rm max}})~dt,\\
A^{\alpha\beta}(k,\omega) &= -\int_0^{t_{\rm max}} {\rm Im} 
[g^{\alpha\beta}(k,t)] \sin(\omega t)~ W(\frac{t}{t_{\rm max}})~dt,
\end{split}
\end{equation}
where $W(x)$ is the Parzen window function, and $t_{\rm max}$ is the 
maximal evolution time (in natural unit). The energy resolution is determined 
by the expression $\varepsilon \simeq 8/t_{\text{max}}$, which corresponds 
to the full width at half maximum (FWHM) of the Fourier transform of 
$W(t/t_{\text{max}})$.
In practical calculations, we perform real-time evolution with retained 
bond dimension up to $D=3000$ on a YC$6\times 15$ lattice under zero 
field, and bond dimension {$D=400$} on a {YC$9\times 18$} lattice 
under $B=2.5$ T in-plane field. The $Y-$type cylindrical lattice utilized in the 
calculations are shown in Fig.~\ref{FigSLattice} below.

% ========== FigS4 ========= %
\begin{figure*}[!htbp]
\includegraphics[angle=0,width=0.9\linewidth]{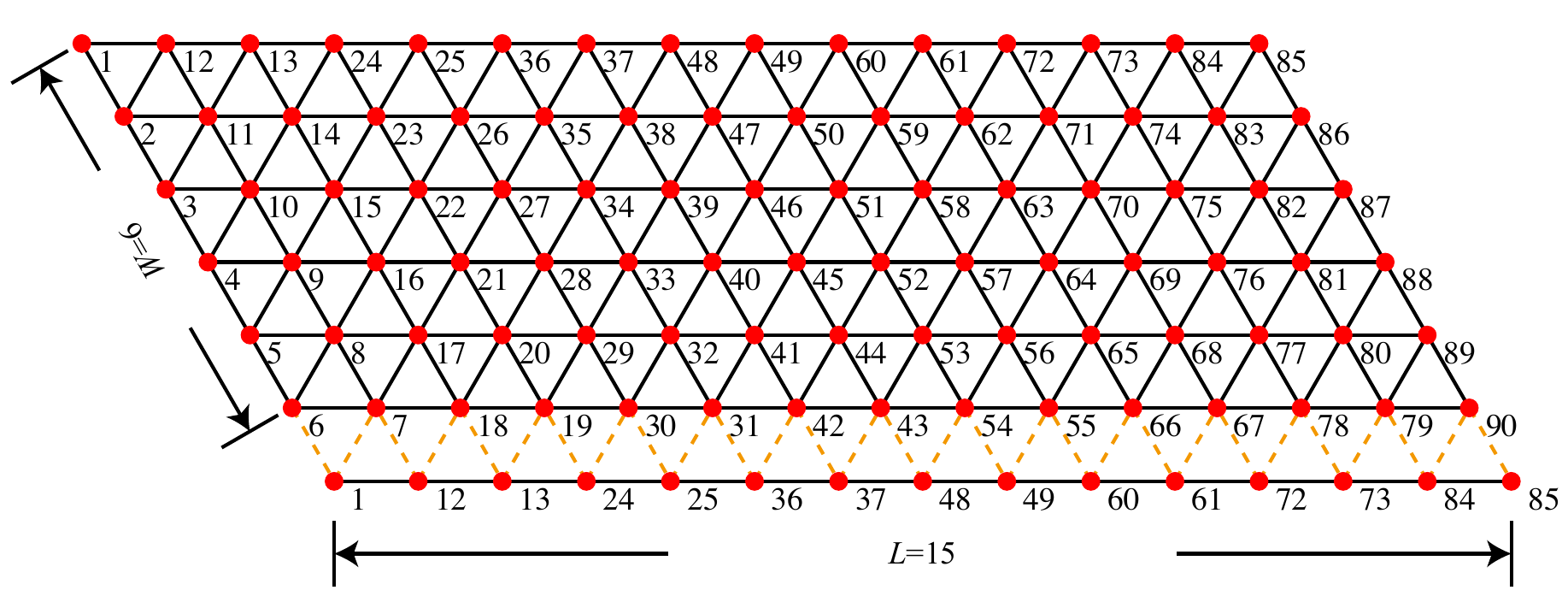}
\caption{The YC$6\times15$ lattice employed in the calculations. 
The number labels the order of the MPS site, and the orange dashed 
lines represent the periodic boundary condition along the $Y$ direction.}
\label{FigSLattice}
\end{figure*}
% ======================== %

\subsection{Data convergence}

In the ground-state calculations, as well as in the time evolutions, 
we retain the same bond dimension $D$, rendering very well converged results. 
Under zero field, we retain up to $D=3000$ bond states for the YC6$\times$15 case 
(see the Y-type cylinder geometry in Fig.~\ref{FigSLattice}); 
while for the case under $B=2.5$~T in-plane field, a relatively small bond dimension 
$D=400$ is sufficient for the accurate simulations of YC$9\times18$ lattice. We 
implement U(1) spin symmetry in our tensor-network calculations using the QSpace 
library~\cite{QSpace,weichselbaum2024QSpace}.

In Figs.~\ref{FigEM_DataConv}(a-d), we show the spectral function $A^{+-, zz}(k={\rm K}, \omega)$ 
obtained with different bond dimensions $D$ and energy resolutions $\varepsilon$. From the results, 
we find the spectral functions are well converged with bond dimension up to $D=3000$, allowing us 
to identify the gapless Goldstone mode at the K point via the extrapolation to $\varepsilon = 0$ limit 
[see Fig.~2(d) of the main text]. In Figs.~\ref{FigEM_DataConv}(e,f), we show the contour plots 
of the calculated spectral functions, obtained with two different bond dimensions $D=1000$ and 2000, 
respectively. Upon increasing the bond dimensions, the magnon-roton excitations exhibit greater clarity, 
and the downward renormalization at the ${\rm M}$ point is also more evident. Therefore, we find $D=2000$ 
suffices to converge the overall profile of spin excitations in the system.

% ========== FigS1 ========= %
\begin{figure}[htbp]
\includegraphics[angle=0,width=0.55\linewidth]{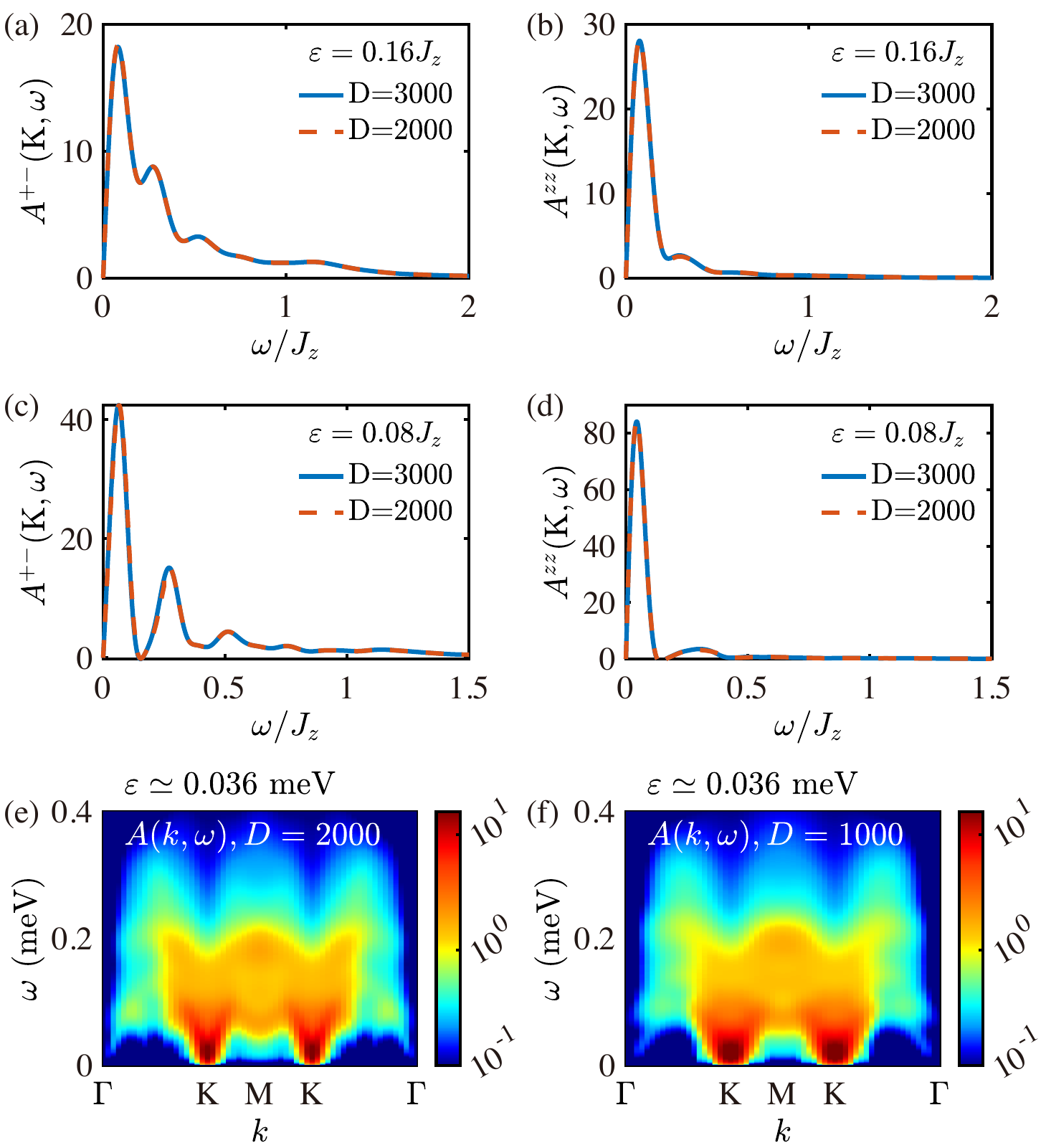}
\caption{(a-d) show the tensor-network results of spectral functions $A^{+-}(k, \omega)$ 
and $A^{zz}(k, \omega)$ calculated at $k={\rm K}$. Results with different bond dimensions 
$D$ and energy resolutions $\varepsilon$ are shown, where $\varepsilon=0.08J_z 
\simeq 0.01$~meV. (e) and (f) show the contour plots of the spectral function $A(k, \omega)$ 
obtained with energy resolution $\varepsilon \simeq 0.036$~meV, where $D=2000$ for 
panel (e) and $D=1000$ for panel (f), respectively.}
\label{FigEM_DataConv}
\end{figure}
% ======================== %

\subsection{Estimation of energy resolution} 
The energy resolution $\varepsilon$ is determined by $W(t/t_{\rm max})$, as the FWHM 
of its Fourier transform. In practice, we choose the Parzen window function as follows
\begin{equation*}
W(t)=
\begin{cases}
-2 (-1+t)^3 & \frac{1}{2}<t\leq 1 \\
 2 (1+t)^3 & -1\leq t<-\frac{1}{2} \\
 1-6 t^2-6 t^3 & -\frac{1}{2}\leq t<0 \\
 1-6 t^2+6 t^3 & 0\leq t\leq \frac{1}{2} \\
 0 & {\rm otherwise}  \\
\end{cases},
\end{equation*}
As the Fourier transformation reads
\begin{equation*}
F[W(t)](\omega) = 6  \sqrt{\frac{2}{\pi}}  \frac{e^{-{\rm i}  \omega} (-1 + e^{{\rm i} 
\omega/2})^4}{\omega^4},
\end{equation*}
and the energy resolution, i.e., FWHM, can be obtained as 
$\varepsilon \simeq 8/t_{\rm max}$.

\subsection{Estimation of the density of states}
The density of states (DOS) can be approximately estimated from the
spectral function. We consider the local spectral density $A(\omega) \equiv \frac{1}{N}
\sum_{k\in {\rm BZ}} A(k,\omega)$, 
where BZ is the first Brillouin zone and $N$ is the system size.
According to Eq.~(\ref{EqS:Akw}), we have
\begin{equation}
\begin{split}
A^{\alpha\beta}(\omega) \equiv \sum_{k\in {\rm BZ}} 
A^{\alpha\beta}(k,\omega) = -\frac{2}{\pi} {\rm Im}  
[\int_{0}^{\infty} e^{{\rm i}\omega t}
{\rm Im}[\frac{1}{N}\sum_k g^{\alpha\beta}(k,t)]].
\end{split}
\label{Awdef}
\end{equation}
Note that 
\begin{equation}
\begin{split}
\frac{1}{N}\sum_k g^{\alpha\beta}(k,t) =& \frac{1}{N}
e^{{\rm i} E_{0} t} 
\sum_{k} \bra{\psi_0}S^{\alpha}_{-k} e^{-{\rm i} H t} 
S^{\beta}_{k} \ket{\psi_0}\\
=&\frac{1}{N} e^{{\rm i} (E_{0}-E_m) t}  
\sum_{k, m}  ||\bra{m}S^{\beta}_k \ket{\psi_0}||^2,
\end{split}
\label{SpecExp}
\end{equation}
where $S^{\beta}_k = \frac{1}{\sqrt{N}}\sum_i e^{{\rm i} k r_i} S_i^{\beta}$, 
$\{\ket{m}\}$ are the eigenstates of $H$ with energy $E_m$ 
and we assume $(S_{-k}^\alpha)^\dagger=S_k^\beta$. Substitue 
Eq.~(\ref{SpecExp}) into Eq.~(\ref{Awdef}), we have
\begin{equation*}
\begin{split}
A^{\alpha\beta}(\omega) =& -\frac{2}{N\pi} \sum_{k,m}  
||\bra{m}S^{\beta}_k \ket{\psi_0}||^2
\int_0^{\infty} {\rm Im} [e^{{\rm i}\omega t} {\rm Im}[e^{{\rm i}(E_0-E_m)t}]]~dt\\
=& \frac{1}{N\pi} \sum_{k,m}  
||\bra{m}S^{\beta}_k \ket{\psi_0}||^2
\int_0^{\infty} {\rm Re}[e^{{\rm i} (\omega + E_0 - E_m) t}] - 
{\rm Re}[e^{{\rm i} (\omega - E_0 + E_m)}]~dt \\
=& \frac{1}{N} \sum_{k,m} ||\bra{m}S^{\beta}_k \ket{\psi_0}||^2 
(\delta(\omega + E_0 - E_m) - \delta(\omega - E_0 + E_m)).
\end{split}
\end{equation*}
Here we introduce the excitation energy $\varepsilon_m \equiv E_m-E_0$ 
and consider only the positive energy part, 
\begin{equation}
A^{\alpha\beta}(\omega > 0) = \frac{1}{N} \sum_{k,m} ||\bra{m}S^{\beta}_k 
\ket{\psi_0}||^2 \, \delta(\omega - \varepsilon_m).
\label{EqAabw}
\end{equation}
Regarding the low-energy excitation states $\ket{m}$ as free gas of 
magnon with energy $\varepsilon_k$, the Hamiltonian 
can be represented as $H \simeq \sum_k \varepsilon_k \gamma_k^\dagger 
\gamma_k$. The excitation states can be represented as $\ket{m} \simeq 
\Pi_{k'}\frac{1}{\sqrt{n_{k'}!}} (\gamma_{k'}^\dagger)^{n_{k'}}\ket{\psi_0}$,
where $\gamma_k$ annihilates a magnon (Bogoliubov quasi-particle). 
Therefore, Eq.~(\ref{EqAabw}) can be rewritten as 
\begin{equation}
A^{\alpha\beta} (\omega > 0) \simeq \frac{1}{N} \sum_{k} 
||\bra{\psi_0}\gamma_{k} S^{\beta}_k \ket{\psi_0}||^2 \, \delta(\omega - \varepsilon_k),
\end{equation}
as multi-magnon excitation states have vanishing contributions. With a rudimentary approximation 
$||\bra{\psi_0}\gamma_{k} S^{\beta}_k \ket{\psi_0}||^2 \sim \mathcal{O}(1)$, we find
\begin{equation}
A^{\alpha\beta} (\omega > 0) \approx \sum_k \delta(\omega - \varepsilon_k)
\end{equation}
to estimate the DOS of magnon excitations with energy $\varepsilon_k$. 

{
Lastly, the local spectral function $A^{\alpha\beta}(\omega)=\frac{1}{N}\sum_k A^{\alpha \beta}(k, \omega)$ can be 
obtained in the real-space calculations, as 
\begin{equation*}
\frac{1}{N}\sum_k g^{\alpha\beta}(k,t) = \frac{1}{N}e^{{\rm i}E_0 t}
\sum_i \bra{\psi_0} S_i^\alpha e^{-{\rm i}Ht} S_i^{\beta}\ket{\psi_0} 
\equiv \frac{1}{N}\sum_i g^{\alpha\beta}_{ii}(t).
\end{equation*}
In practice, we choose to compute the local spectral density at site $i$ in the center of the lattice, to 
facilitate the calculations and reduce the finite-size effect.
}

With this, the low-temperature entropy and specific heat can be computed as 
\begin{equation}
\begin{split}
\tilde{S}(T) =& \int_0^\infty A(\omega) [\frac{\omega/T}{e^{\omega/T}-1}-
{\rm ln}(1-e^{-\omega/T})]~d\omega, \\
\tilde{C}(T) =& T\frac{\partial \tilde{S}(T)}{\partial T},
\end{split}
\end{equation}
where $A(\omega) \equiv A^{+-}(\omega) + A^{zz}(\omega)$ is normalized such 
that $\int_0^\infty A(\omega)~d\omega= 1$. 

\subsection{Magnon-maxon-roton excitations}
Here we show the  magnon-maxon-roton fitting of the calculated excitation 
dispersion. The terminology ``maxon" and ``roton" is derived from the excitations 
spectrum of superfluid helium-4, as referenced in \cite{Fukushima1989}. 
The magnon has a gapless linear excitation, also known as the Goldstone mode 
arising from U(1) symmetry breaking. The maxon represents the quadratic excitation 
near the round peak of the dispersion curve, whereas the roton signifies the quadratic 
excitation at the minimum. We fit the numerical results with the undetermined function 
\begin{equation}
\label{EqS:Fitting}
f(\xi)=
\begin{cases}
-k_0(\xi-1/3) + \Delta_0, & \xi \leq 1/3 \\
k_0(\xi-1/3) + \Delta_0, & 1/3 < \xi \leq \xi_0 \\
-a_{\rm M} (\xi-\xi_{\rm M})^2 + \Delta_{\rm M}, & \xi_0 < \xi \leq \xi_1\\
a_{\rm R} (\xi-1/2)^2 + \Delta_{\rm R}, & \xi_1 < \xi \\
\end{cases}
\end{equation}
where $k_0, \Delta_0, \xi_0, a_{\rm M}, \xi_{\rm M}, \Delta_{\rm M}, \xi_1, a_{\rm R}, 
\Delta_{\rm R}$ are the fitting parameters. The first two lines are the 
Goldstone (or pseudo-Goldstone) mode, the third line is the maxon part, 
and the last line is the roton dispersion. Note that $f(\xi)$ should be a continuous 
function, thus the fitting parameters can be reduced to $k_0, \Delta_0, 
\xi_0,  \xi_{\rm M}, \xi_1, \Delta_{\rm R}$. The corresponding fitting results 
are shown in Fig.~\ref{FigSFitting}, where we find the magnon-maxon-roton 
dispersion well describes the magnetic excitations from both $A^{zz}$ in the
(a) panel and $A^{+-}$ in panel (b).

In the study of Goldstone mode excitations derived from $A^{zz}(k, \omega)$, 
we presume $\Delta_0$ to be zero, which is in line with the expected U(1) symmetry 
breaking scenario. On the contrary, when assessing the $A^{+-}(k, \omega)$ 
excitation channel, we permit $\Delta_0$ to be a free parameter that is accurately 
adjusted to align with the data.

% ========== FigS8 ========= %
\begin{figure*}[!htbp]
\includegraphics[angle=0,width=0.7\linewidth]{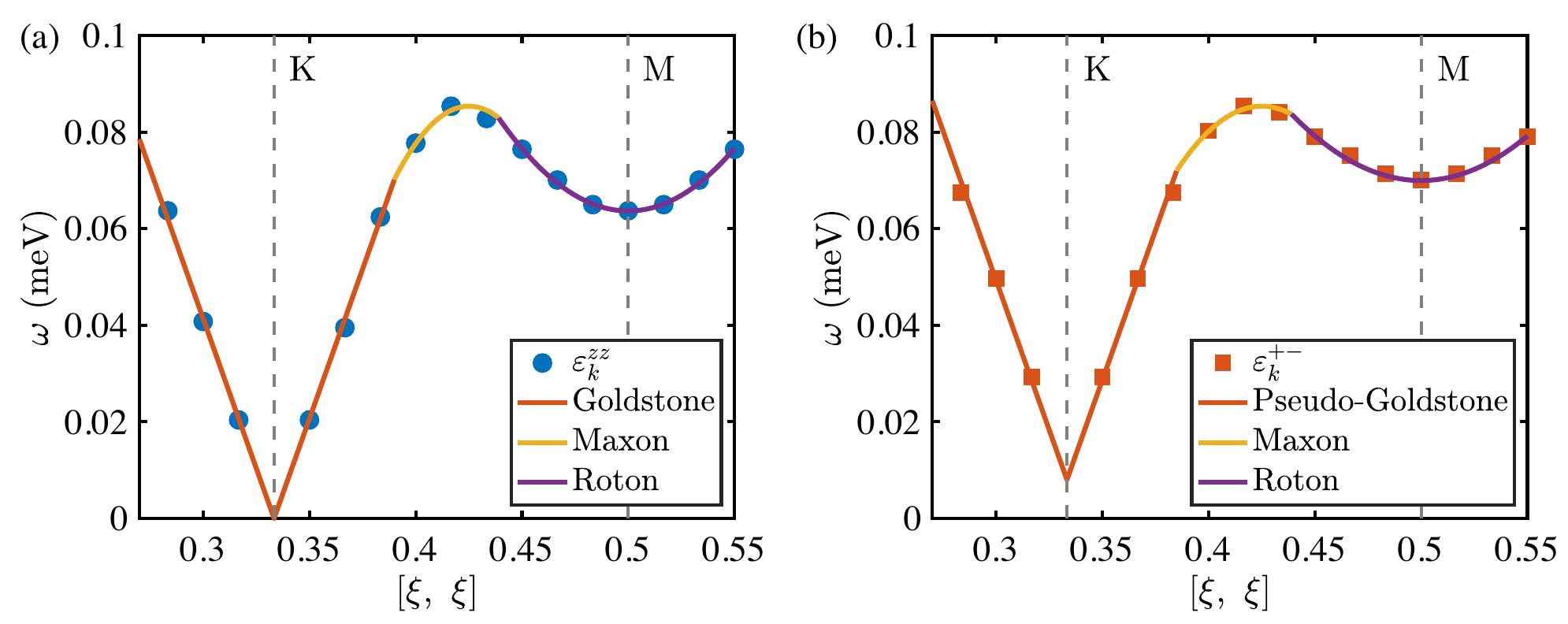}
\caption{The magnon-maxon-roton dispersion $\varepsilon_k$. 
Based on the data $\varepsilon_k^{zz}$ derived from $A^{zz}(k, \omega)$, 
we arrive at the fitting parameters $k_0=1.24, \Delta_0=0, \xi_0=0.39, 
\xi_{\rm M}=0.439, \xi_1=0.425, \Delta_{\rm R}=0.0637$. 
For $\varepsilon_k^{+-}$ from $A^{+-}(k, \omega)$, we obtain the parameters 
$k_0=1.24, \Delta_0=0.008, \xi_0=0.385,  \xi_{\rm M}=0.439, \xi_1=0.425, 
\Delta_{\rm R}=0.07$. See definitions of these fitting parameters in 
Eq.~(\ref{EqS:Fitting}).}
\label{FigSFitting}
\end{figure*}
% ======================== %

\subsection{Spin excitation continuum}
Recently, there has been intense research interest on the spin excitation continuum in the easy-axis 
TLAF systems. Here we provide accurate model calculations with sufficiently large bond dimension 
to address this question. In Fig.~\ref{FigHen}(a), we present the experimental results; while in 
Figs.~\ref{FigHen}(b-f), we calculate the spectral functions $A^{+-}(k, \omega)$ with varying energy 
resolutions $\varepsilon$. Notably, the $W$-shaped excitation continuum remains broadened in our
tensor-network calculations even as the energy resolution is increased.

The origin of the excitation continuum may be attributed to, among others, two possible sources: 
magnon-magnon interactions and the proximity to a QSL phase. In the TLAF systems, magnon 
interactions can induce magnon decay, e.g., into two-magnon continuum, thereby generating 
a broad spectrum of excitations. Another possibility is that the triangular-lattice spin supersolid 
may be positioned close to certain QSL phase~\cite{chen2024continuum,sheng2024continuum}, 
which could give rise to a multi-spinon continuum in the relatively high-energy regime.

% ========== FigS5 ========= %
\begin{figure}[]
\includegraphics[angle=0,width=0.55\linewidth]{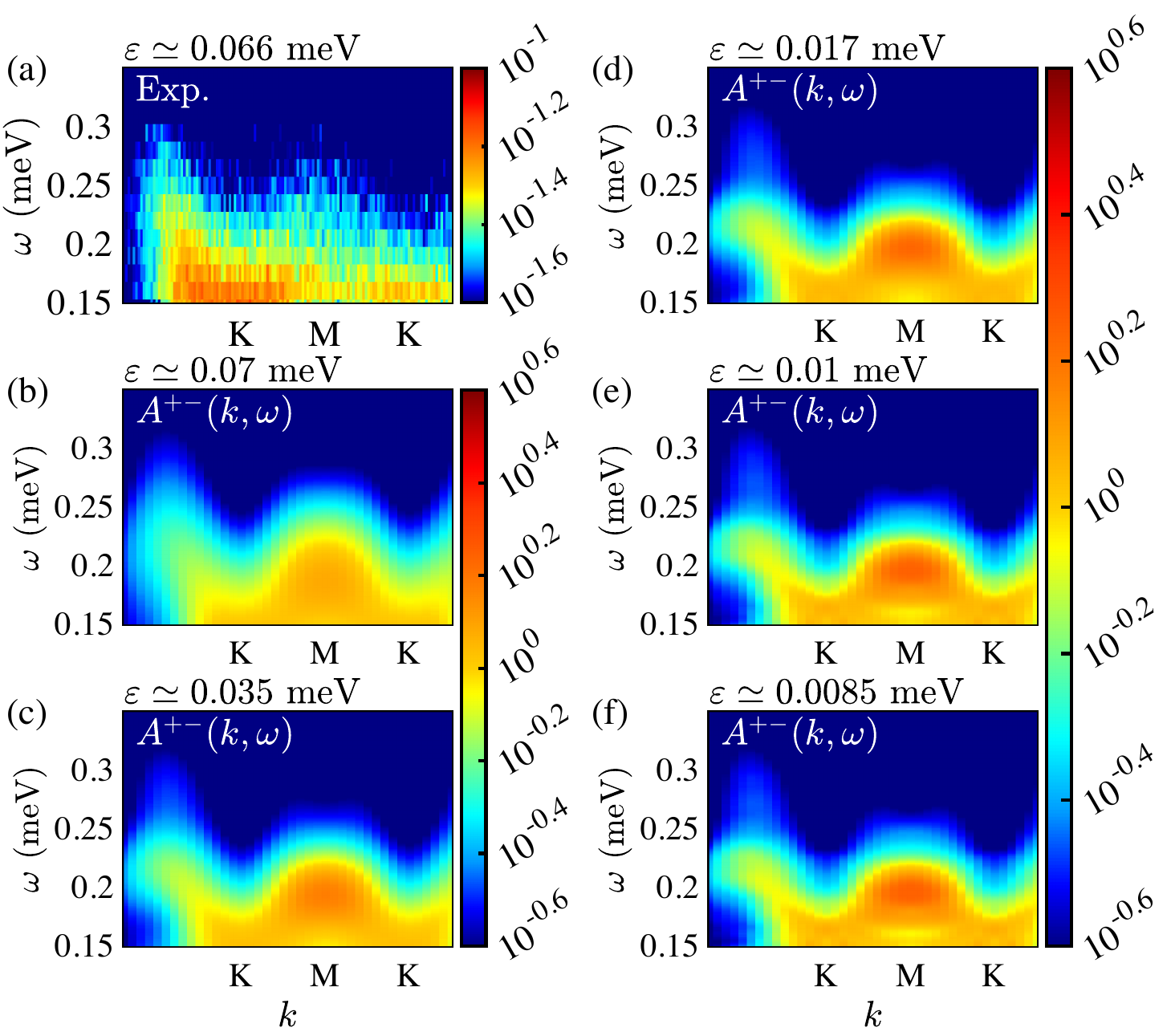}
\caption{Spin excitation continuum in the relatively high-energy regime $\omega \gtrsim J_z$. 
(a) shows the contour plot of the INS measurements, and (b-f) are the calculated spectral functions 
$A^{+-}(k,\omega)$ with different resolutions $\varepsilon$ (down to $0.0085$~meV). We retain 
bond dimension $D=1000$ in the simulations, which generates well converged dynamical correlations 
$A(k,\omega)$ in the concerned energy regime $\omega \geq 0.15$~meV (see Fig.~\ref{FigEM_DataConv}).}
\label{FigHen}
\end{figure}
% ======================== %

% ======================== %
\section{Linear spin-wave calculations} 

We detail the derivations of linear spin-wave {(LSW)} theory for the easy-axis 
triangular-lattice antiferromagnetic (TLAF) model
$$H = \sum_{\langle i,j\rangle} J_{xy} (S_i^xS_j^x+S_i^yS_j^y) + 
J_z S_i^zS_j^z,$$ where the anisotropy parameter $\Delta = J_z/J_{xy}$. 
The ground state of the easy-axis TLAF model with $\Delta > 1$ is a 
Y-shaped state in the $x$-$z$ plane. There are three sublattices, namely $A$, 
$B$, and $C$, and thus three kinds of Holstein-Primakoff bosons, 
$a$, $b$, and $c$ are introduced. For the sublattice $A$, we have
\begin{equation*}
\begin{split}
	& S^z = (S - a^\dagger a),\\ 
	& S^x = \frac{\sqrt{2S}}{2} (a + a^\dagger),\\
	& S^y = \frac{\sqrt{2S}}{2{\rm i}} (a - a^\dagger).
\end{split}
\end{equation*} 
On the other two sublattices, there are angles $\pm \theta$ between spins 
on A and B(C) sublattices. For sublattice $B$, the transformation reads
\begin{equation*}
\begin{split}
	&S^z=\cos\theta~ (S-b^\dagger b)- \sin \theta~\frac{\sqrt{2S}}{2}(b+b^\dagger),\\
	&S^x=\sin\theta~ (S-b^\dagger b)+\cos \theta~\frac{\sqrt{2S}}{2}(b+b^\dagger),\\
	&S^y=\frac{\sqrt{2S}}{2{\rm i}}(b-b^\dagger),
\end{split}
\end{equation*}
and for sublattice $C$
\begin{equation*}
\begin{split}
	&S^z=\cos\theta~ (S-c^\dagger c)+ \sin \theta~\frac{\sqrt{2S}}{2}(c+c^\dagger), \\
	&S^x=-\sin\theta~ (S-c^\dagger c)+\cos \theta~\frac{\sqrt{2S}}{2}(c+c^\dagger),\\
	&S^y=\frac{\sqrt{2S}}{2{\rm i}}(c-c^\dagger).
\end{split}
\end{equation*}

Through the Holstein-Primakoff and Fourier transformations, 
we arrive at a quadratic form of the Hamiltionian $$H_{\rm HP} 
= \sum_k \alpha_k^\dagger H_0(k) \alpha_k,$$ where 
$\alpha_k^\dagger  = (a_k^\dagger~b_k^\dagger ~c_k^\dagger
~a_{-k}~b_{-k}~c_{-k})$ denote the magnon creation operators. 
The quadratic Hamiltonian is $H_0(i,j)= S \, A_0(i,j) \, Z(i,j)$, 
where the $k$-independent symmetric part is $A_0=A + A^{\rm T}$, 
$i,j$ label the matrix index, and $S=1/2$ for the present case. 
$A$ is a $6\times6$ upper triangular matrix
\begin{equation*}
A=
\begin{pmatrix}
M & N \\
0 & M
\end{pmatrix},
\end{equation*}
with 
\begin{equation*}
M=
\begin{pmatrix} 
-3\Delta  \cos\theta 
& \frac{\cos \theta+1}{2}
& \frac{\cos \theta+1}{2}\\
0
& \frac{3}{2}(\sin^2\theta-\Delta \cos^2\theta - \Delta \cos\theta)
& \frac{\cos^2\theta +1-\Delta \sin^2\theta}{2}\\
0
& 0
& \frac{3}{2}(\sin^2\theta-\Delta \cos^2\theta - \Delta \cos\theta)
\\
\end{pmatrix},
\end{equation*}
and
\begin{equation*}
N=
\begin{pmatrix}
0
& \frac{\cos \theta-1}{2}
& \frac{\cos \theta-1}{2} \\
\frac{\cos \theta-1}{2}
& 0
& \frac{\cos^2\theta -1-\Delta \sin^2\theta}{2}\\
 \frac{\cos \theta-1}{2}
& \frac{\cos^2\theta -1-\Delta \sin^2\theta}{2}
& 0 
\end{pmatrix},
\end{equation*}
and $Z(k)$ is a $6\times 6$ matrix 
\begin{equation*}
Z(k) = 
\begin{pmatrix}
1 & z & z^*& 1 & z & z^*\\
z^*& 1 & z & z^*& 1 & z \\
z & z^*& 1 & z & z^*& 1 \\
1 & z & z^*& 1 & z & z^*\\
z^*& 1 & z & z^*& 1 & z \\
 z & z^*& 1 & z & z^*& 1\\
\end{pmatrix},
\end{equation*}
with $z=\sum_i e^{{\rm i} k \delta_i}$ with $\delta_1 = (1,0)$, $\delta_2 = 
(-\frac{1}{2}, \frac{\sqrt{3}}{2})$, $\delta_3 = (-\frac{1}{2}, -\frac{\sqrt{3}}{2})$.
The angle $\theta$ can be obtained by minimizing the ground-state energy 
$E_0 = S^2(2\Delta \cos \theta  + \Delta \cos^2\theta  - \sin^2 \theta )$,
and we find $\theta = 128.91^\circ$ for the realistic parameter $\Delta = 1.68$.
With this, we diagonalize {the Hamiltonian by Bogoliubov transformation 
$\gamma_k = U_k \alpha_k$ as $H_{\rm HP} = \sum_k \gamma_k^\dagger 
D(k) \gamma_k$}  and obtain the linear spin-wave 
dispersion shown in {Fig.~3} of the main text.

Besides, the dynamical spin structure can also be obtained by LSW calculations following as
\begin{equation}
S^{\alpha \alpha}(k,\omega) \equiv 2 \pi \sum_m |\bra{m} S^\alpha_k \ket{0}|^2
 \delta(\omega-E_m + E_0),
\end{equation}
where $\ket{m}$ denotes the ground state for $m = 0$ or an excited 
state for $m \neq 0$, and $E_m$ is the corresponding energy eigenvalue. 
Here we ignore the constant number and only consider one-magnon excited 
state, i.e. $\ket{m} = \gamma_k^{i \dagger} \ket{0}$ denotes the component of the vector. 
Thus we have 
\begin{equation}
\begin{split}
S^{\alpha \alpha}(k,\omega) &= \sum_{m, i} |\bra{0} \gamma_k^i S^\alpha_k \ket{0}|^2
 \delta(\omega-E_m + E_0)\\
 &= \sum_{m, i, j} |\bra{0} U_k^{i,j} \alpha_k^j S^\alpha_k \ket{0}|^2
 \delta(\omega-E_m + E_0).
\end{split}
\end{equation}
By replacing the delta function with the Parzen window function, we introduce the 
energy resolution $\varepsilon$ and show the corresponding results in Fig.~\ref{FigSW}. 
From the 0~T results, we find the gapless Goldstone mode can be captured, whereas the 
pseudo-Goldstone gap and ${\rm M}$-point roton minima are not reproduced within the 
LSW framework. In sharp contrast, for the case with 2.5~T in-plane field, LSW theory 
works well and provide very accurate spin excitation results instead.

% ======== FigS2 ======== %
\begin{figure}[htbp]
\includegraphics[angle=0,width=0.55\linewidth]{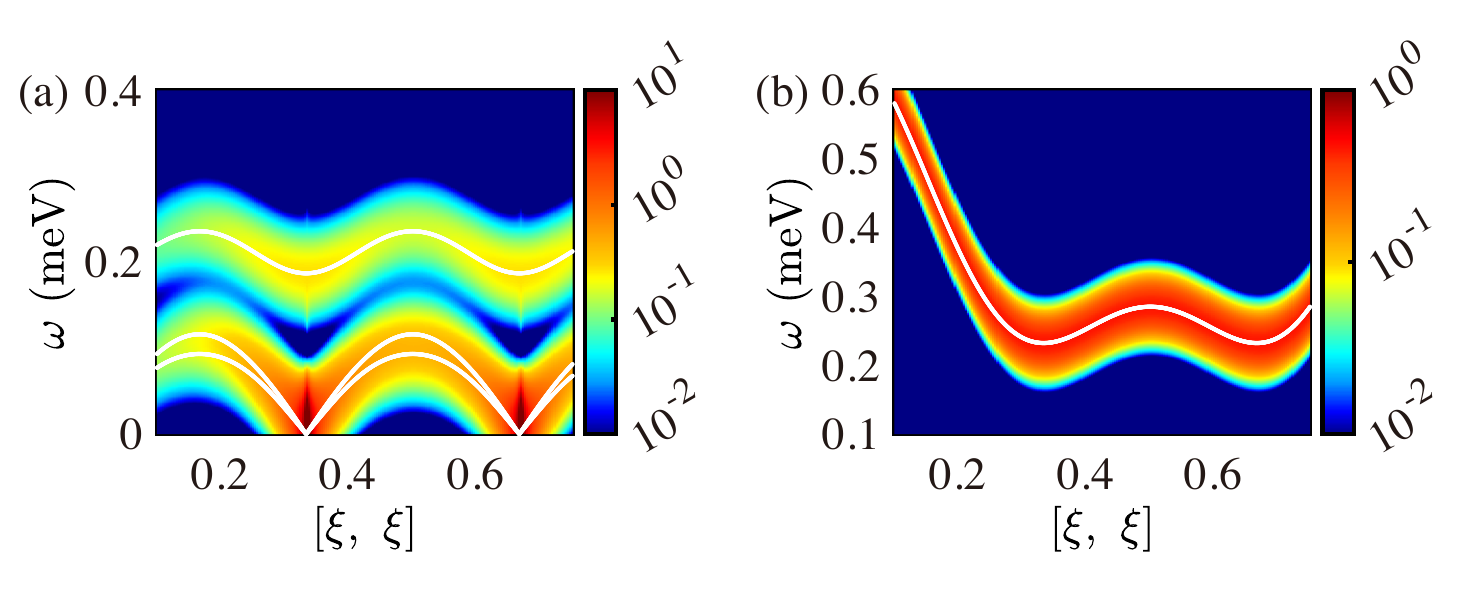}
\caption{The LSW results of dynamical spin structure factor under (a) zero field and 
(b) 2.5 T in-plane field. An energy resolution of $\varepsilon \simeq 0.066$~meV is 
selected to facilitate direct comparison with experimental data.}
\label{FigSW}
\end{figure}
% ======================== %

\end{document}